\documentclass[twocolumn]{rbef}
\usepackage[T1]{fontenc}
\usepackage{lipsum}
\usepackage{listings}
\usepackage{hyperref}
\usepackage{url}
\usepackage{color}
\usepackage{algorithm}
\usepackage[noend]{algpseudocode}
\usepackage{amsmath}
\usepackage{amssymb}
\usepackage{cite}
\usepackage[normalem]{ulem}
\usepackage{flushend}
\usepackage{cuted}

\DeclareOldFontCommand{\rm}{\normalfont\rmfamily}{\mathrm}
\DeclareOldFontCommand{\sf}{\normalfont\sffamily}{\mathsf}
\DeclareOldFontCommand{\tt}{\normalfont\ttfamily}{\mathtt}
\DeclareOldFontCommand{\bf}{\normalfont\bfseries}{\mathbf}
\DeclareOldFontCommand{\it}{\normalfont\itshape}{\mathit}
\DeclareOldFontCommand{\sl}{\normalfont\slshape}{\@nomath\sl}
\DeclareOldFontCommand{\sc}{\normalfont\scshape}{\@nomath\sc}

\titulocabecalho{GC espectral}
\autorcabecalho{Lima et al.}

\numeracao{e20200007}
\volume{42}
\numero{xx}
\ano{2020}
\doi{http://dx.doi.org/10.1590/1806-9126-RBEF-2020-0007}
\tipodeartigo{Artigos Gerais}

\titulo{Granger causality in the frequency domain: derivation and applications}

\definecolor{mauricio}{RGB}{0,0,255}

\newcommand*{\mycomment}[1]{{\tiny \# #1}}

\begin{document}
	
	\author[1,2]{Vinicius Lima}
	\author[1,2]{Fernanda Jaiara Dellajustina}
	\author[1]{Renan O. Shimoura}
	\author[1]{Mauricio Girardi-Schappo}
	\author[1]{Nilton L. Kamiji}
	\author[1]{Rodrigo F. O. Pena}
    \author[1]{Antonio C. Roque}
	\affil[1]{Laborat\'orio de Sistemas Neurais, Departamento de Física,  Faculdade de Filosofia, Ciências e Letras de Ribeir\~ao Preto, Universidade de S\~ao Paulo, Ribeir\~ao Preto, Brasil}
	\affil[2]{Esses autores contribuíram igualmente na escrita e elaboração do presente trabalho.}\thanks{\href{emailto:vinicius.lima.cordeiro@gmail.com}{vinicius.lima.cordeiro@gmail.com}}\thanks{\href{emailto:antonior@usp.br}{antonior@usp.br}}

\begin{primeirapagina}


	\begin{abstract}
Físicas e físicos têm começado a trabalhar em áreas onde é necessária a análise de sinais
ruidosos. Nessas áreas, tais como a Economia, a Neurociência e a Física, a noção de causalidade deve ser interpretada como uma medida estatística.
Introduzimos ao leitor leigo a causalidade de Granger entre duas séries temporais e ilustramos como calculá-la:
um sinal $X$ ``Granger-causa'' $Y$ se a observação do passado de $X$ aumenta a previsibilidade do futuro de $Y$ em comparação com o que é possível prever apenas pela observação do passado de $Y$.
Em outras palavras, para haver causalidade de Granger entre dois sinais basta que a informação do passado de um melhore a previsão do futuro de outro,
mesmo na ausência de mecanismos físicos de interação.
Apresentamos a derivação da causalidade de Granger nos domínios do tempo e da frequência e damos
exemplos numéricos através de um método não-paramétrico no domínio da frequência.
Métodos paramétricos são abordados no Apêndice.
Discutimos limitações e aplicações desse método e outras alternativas para medir causalidade.
    \palavraschave{Causalidade de Granger, processo autoregressivo, causalidade de Granger condicional, estimação não-paramétrica}

	\end{abstract}

	\begin{otherlanguage}{english}

	\begin{abstract}

Physicists are starting to work in areas where noisy signal analysis is required.
In these fields, such as Economics, Neuroscience, and Physics, the notion of causality
should be interpreted as a statistical measure.
We introduce to the lay reader the Granger causality between two time series
and illustrate ways of calculating it:
a signal $X$ ``Granger-causes'' a signal $Y$ if the observation of the past of $X$ increases the predictability of the future of $Y$ when compared to the same prediction done with the past of $Y$ alone. In other words, for Granger causality between two quantities it suffices that information extracted from the past of one of them improves the forecast of the future of the other, even in the absence of any physical mechanism of interaction.
We present derivations of the Granger causality measure in the time and frequency domains and give numerical examples using a non-parametric estimation method in the frequency domain. Parametric methods are addressed in the Appendix.
We discuss the limitations and applications of this method and other alternatives to measure causality.
	\end{abstract}
	\end{otherlanguage}

	\end{primeirapagina}
\saythanks

\section{Introduction}\label{codigoneural}

The notion of causality has been the concern of thinkers at least since the ancient Greeks~\cite{falcon2019aristotle}. More recently, Clive Granger~\cite{granger1969investigating}, in his paper entitled ``Investigating Causal Relations by Econometric Models and Cross-spectral Methods'' from 1969, elaborated
a mathematical framework to describe a form of causality -- henceforth called
\textit{Granger Causality}\footnote{It is also referred as Wiener-Granger causality.} (GC) in order to distinguish it from
other definitions of causality. Given two stochastic variables, $X(t)$ and $Y(t)$, there is
a causal relationship (in the sense of Granger) between them if the past observations of $Y$ help to predict the current state of $X$, and \textit{vice-versa}. If so,
then we say that $Y$ Granger-causes $X$.
Granger was inspired by the definition of causality from Norbert Wiener~\cite{wiener1956theory}, in which $Y$ causes $X$ if knowing the past
of $Y$ increases the efficacy of the prediction of the current state of $X(t)$ when compared to the prediction of $X(t)$ by the past values of $X$ alone\footnote{Other notions of causality have been defined, one worth mentioning is Pearl's causality \cite{pearl1995causal}. Over the years, Pearl's causality has been revised by him and colleagues in a series of published works \cite{halpern2005causes,halpern2015modification}.}. 

In the multidisciplinary science era, more and more physicists are involved in research in other areas, such as Economics and Neuroscience. These areas usually have big data
sets. Data analysis tools, such as GC, come in handy to extract meaningful knowledge from these sets.

Causality inference via GC has been widely applied in different areas of science, such as:
prediction of financial time series~\cite{vyrost2015granger, candelon2016nonparametric, ding2016crude},
earth systems~\cite{runge2019earth}, atmospheric systems~\cite{smirnov2009clima},
solar indices~\cite{amblard2012review},
turbulence~\cite{amblard2012review,tissot2014turbo},
inference of information flow in the brain of different
animals~\cite{brovelli2004beta,ding200617,matias2014modeling, hu2015comparison,seth2015granger},
and inference of functional networks of the
brain using fMRI~\cite{havlicek2010dynamic, liao2010evaluating},
MEG~\cite{pollonini2010functional}, and
EEG~\cite{barrett2012granger}.
It appears as an alternative to measures like
linear correlations~\cite{lynall2010functional}, mutual information~\cite{jin2011abnormal, lima2019information}, partial directed coherence~\cite{baccala2001partial}, ordinary coherence~\cite{fries2005}, directed transfer function~\cite{kaminski1991newmethod}, spectral coherence~\cite{la2014human}, and transfer entropy~\cite{schreiber2000measuring, vicente2011transfer}, being usually easier to calculate
since it does not rely on the estimation of
probability density functions of one or more variables.

\begin{figure}[bp]
    \centering
    \includegraphics[scale=0.16]{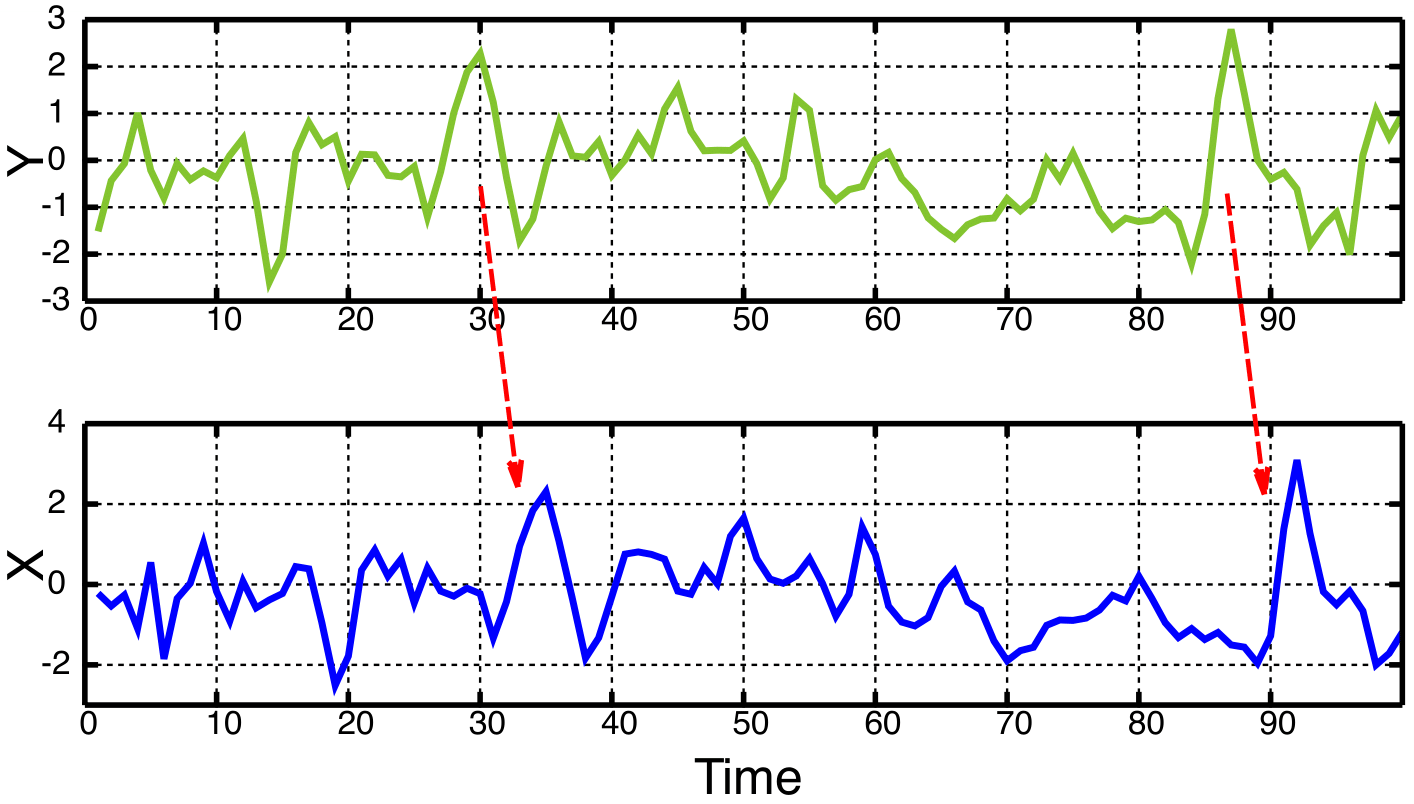}
    \caption{When time series $Y(t)$ Granger-causes time series $X(t)$, the patterns in $Y(t)$ are approximately repeated in $X(t)$ after some time lag (two examples are indicated with arrows). Thus, past values of $X$ can be used for the prediction of future values of $Y$.}
    \label{fig:Figura1}
\end{figure}

The definition of GC involves the prediction of future values of stochastic time series
(see Fig.~\ref{fig:Figura1}).
The measurement of the GC between variables may be done in both the time and the frequency
domains~\cite{geweke1982measurement, geweke1984measures,baccala2001partial,dhamala2008estimating}.

In the present work, we will focus on the frequency domain representation of the GC \cite{geweke1982measurement,baccala2001partial} and, for pedagogical purposes, will discuss illustrative examples from previous works by other authors~\cite{dhamala2008estimating,baccala2001partial}. Our main goal is to provide a basic notion of the GC measure to a reader not yet introduced to this subject.

This work is organized as follows: in
Section~\ref{sec::ar_models}, we present the concept of an autoregressive process -- a model of
linear regression in which GC is based (it is also possible to formulate GC for nonlinear systems, however such a formulation results in a more complex analysis which is beyond the scope of this work \cite{seth2007granger, marinazzo2011nonlinear}). Sections~\ref{sec:gc_time} and~\ref{sec:gc_freq}
are used to develop the mathematical concepts and definitions of
the GC both in the time and frequency domains. In Section~\ref{sec:non_parametric}, we introduce the nonparametric method to estimate GC through
Fourier and wavelet transforms \cite{dhamala2008estimating}. In Section~\ref{sec:cgc} we introduce examples of the conditional GC (cGC) to determine known 
links between the elements of a simple network.
We then close the paper by discussing applications, implications
and limitations of the method.

\section{Autoregresive process}
\label{sec::ar_models}

\begin{figure}[!tp]
    \centering
    \includegraphics[scale=0.2]{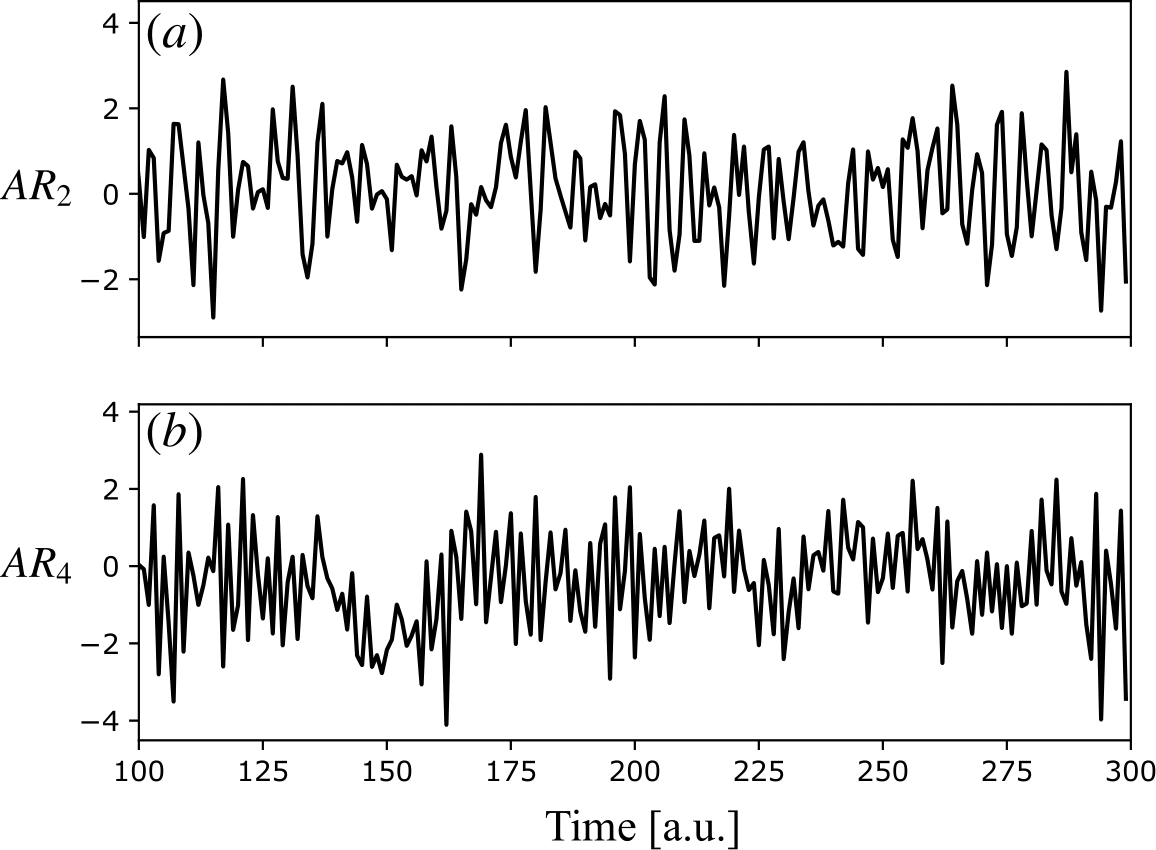}
    \caption{Autoregressive processes. (a) time series of an $AR_{2}$ process with coefficients
    $(a_1, a_2)=(0.3, -0.5)$. (b) time series of an $AR_{4}$ process with coefficients $(a_1, a_2, a_3, a_4)=(-0.2, 0.5, 0.6, -0.2)$.}
    \label{fig:Figura2}
\end{figure}

Autoregressive processes form the basis for the parametric estimation of the GC, so in this section we introduce the reader to the basic concepts of such processes~\cite{boxjenkins2015time}.
A process $X(t)$ is autoregressive of order $n$ (\textit{i.e.}, $AR(n)$) if its state at time $t$
is a function of its $n$ past states:
\begin{equation}
    X(t) = \sum_{i=1}^{n}a_iX(t-i) + \epsilon(t),
    \label{eq:ar}
\end{equation}
where $t$ is the integer time step, and the real coefficients $a_i$ indicate the weighted contribution from
$i$ steps in the past, to the current state $t$ of $X$. The term $\epsilon(t)$ is a noise source with variance
$\Sigma$ that models any external additive contribution to the determination of 
$X(t)$. If $\Sigma$ is large, then the process is weakly dependent on its past states
and $X(t)$ may be regarded as just noise. 
Fig.~\ref{fig:Figura2} shows examples of an $AR(2)$ (a) and an $AR(4)$ (b).

Fitting the autoregressive coefficients $a_{i}$ and the noise variance $\Sigma$, for a recorded signal, is usually done by solving a Yule-Walker set of
equations~\cite{eshel2003yule, ding200617}. For a brief review on this topic see the Section~A of the Appendix.

\section{Granger causality in time domain}
\label{sec:gc_time}

In this section we develop the mathematical concepts and definitions of GC in time domain.
Consider two stochastic signals, $X_1(t)$ and $X_2(t)$. We assume that these signals may be modeled by
autoregressive stochastic processes of order $n$, independent of each other,
such that their states in time $t$ could be estimated by their $n$ past values:
\begin{align}
    X_{1}(t) = \sum_{i=1}^{n}a_iX_{1}(t-i) + \epsilon_{1}(t),
    \label{eq:ar_x}\\
    X_{2}(t) = \sum_{i=1}^{n}c_iX_{2}(t-i) + \epsilon_{2}(t),
    \label{eq:ar_y}
\end{align}
where the variances of $\epsilon_1$ and $\epsilon_2$ are, respectively,
$\Sigma_{11}$ and $\Sigma_{22}$, and
the coefficients $a_i$ and $c_i$ are adjusted in order to minimize $\Sigma_{11}$
and $\Sigma_{22}$.

However, we may also assume that the signals $X_1(t)$ and $X_2(t)$ are each modeled
by a combination of one another, yielding
\begin{align}
    X_{1}(t) = \sum_{i=1}^{n}a_iX_{1}(t-i) + \sum_{i=1}^{n}b_iX_{2}(t-i) + \epsilon^{*}_{1}(t),
    \label{eq:ar_xy}\\
    X_{2}(t) = \sum_{i=1}^{n}c_iX_{2}(t-i) + \sum_{i=1}^{n}d_iX_{1}(t-i) + \epsilon^{*}_{2}(t),
    \label{eq:ar_yx}
\end{align}
where the covariance matrix is given by
\begin{equation}
    \pmb\Sigma = \begin{bmatrix} 
                    \Sigma^{*}_{11} & \Sigma^{*}_{12} \\
                    \Sigma^{*}_{21} & \Sigma^{*}_{22} 
                 \end{bmatrix}.
    \label{eq:cov_mat}
\end{equation}

Here, $\Sigma^{*}_{11}$, $\Sigma^{*}_{22}$ are the variances of $\epsilon^{*}_{1}$ and $\epsilon^{*}_{2}$ respectively,
and $\Sigma^{*}_{12} = \Sigma^{*}_{21}$ is the covariance of $\epsilon^{*}_{1}$ and $\epsilon^{*}_{2}$.
Again, the coefficients $a_i$, $b_i$, $c_i$ and $d_i$ are adjusted to minimize
the variances $\Sigma^*_{11}$ and $\Sigma^*_{22}$.

If $\Sigma^{*}_{11} < \Sigma_{11}$, then the addition of $X_2(t)$ to $X_1(t)$ generated a better fit
to $X_1(t)$, and thus enhanced its
predictability. In this sense,
we may say there is a causal relation from $X_2$ to $X_1$, or simply that
$X_2$ Granger-causes $X_1$. The same applies for the other signal:
if $\Sigma^{*}_{22} < \Sigma_{22}$, then $X_1$ Granger-causes $X_2$ because
adding $X_1$ to the dynamics of $X_2$ enhanced its predictability.

We may summarize this concept into the definition of the \textit{total causality index},
given by
\begin{equation}
    F_{1.2} = \log\left(\dfrac{\Sigma_{11}\Sigma_{22}}{\det(\pmb\Sigma)}\right) = \log\left(\dfrac{\Sigma_{11}\Sigma_{22}}{\Sigma^{*}_{11}\Sigma^{*}_{22}-(\Sigma^{*}_{12})^2}\right).
    \label{eq:causalidade_total}
\end{equation}
If $F_{1.2}>0$, there is some Granger-causal relation between $X_1$ and $X_2$, because
either $\Sigma^{*}_{11}<\Sigma_{11}$ or $\Sigma^{*}_{22}<\Sigma_{22}$, otherwise
there is correlation between $X_1$ and $X_2$ due to $\Sigma^{*}_{12}>0$.
If neither Granger-causality nor correlations are present, then $F_{1.2}=0$.

To know specifically whether there is Granger causality from 1 to 2 or from 2 to 1, we may use
the specific indices:
\begin{align}
F_{1\rightarrow 2} = \log\left(\dfrac{\Sigma_{22}}{\Sigma_{22}^{*}}\right),
\label{eq:x2y}\\
F_{2\rightarrow 1} = \log\left(\dfrac{\Sigma_{11}}{\Sigma^{*}_{11}}\right),
\label{eq:y2x}\\
F_{1\leftrightarrow 2} = \log\left(\dfrac{\Sigma^{*}_{11}\Sigma^{*}_{22}}{\det(\pmb\Sigma)}\right),
\label{eq:xy}
\end{align}
such that
\begin{equation}
F_{1.2} = F_{1\rightarrow 2} + F_{2\rightarrow 1} + F_{1\leftrightarrow 2},
\label{eq:causalidade_total2}
\end{equation}
where $ F_{1\rightarrow 2}$ defines the causality from $X_{1}(t)$ to $X_{2}(t)$,
$F_{2\rightarrow 1}$ is the causality from $X_{2}(t)$ to $X_{1}(t)$,
and $F_{1\leftrightarrow 2}$ is called \textit{instantaneous causality} due to correlations
between $\epsilon^*_1$ and $\epsilon^*_2$. Just as for the total causality case,
these specific indices are greater than zero if there is Granger causality, or zero otherwise.

\section{Granger causality in frequency domain}\label{sec:gc_freq}

In order to derive the GC in frequency domain, we first define the lag operator $L^k$, such that
\begin{equation}
    \label{eq:delayop}
    L^{k}X(t)=X(t-k),
\end{equation}
delays $X(t)$ by $k$ time steps, yielding $X(t-k)$. We may then rewrite
equations~\eqref{eq:ar_xy} and~\eqref{eq:ar_yx} as:
\begin{align}
    X_{1}(t) = \left(\sum_{i=1}^{n}a_iL^{i}\right)X_{1}(t) + \left(\sum_{i=1}^{n}b_iL^{i}\right)X_{2}(t) + \epsilon^{*}_{1}(t),
    \label{eq:ar_xy2}\\
    X_{2}(t) = \left(\sum_{i=1}^{n}c_iL^{i}\right)X_{1}(t) + \left(\sum_{i=1}^{n}d_iL^{i}\right)X_{2}(t) + \epsilon^{*}_{2}(t),
    \label{eq:ar_yx2}
\end{align}
and rearrange their terms to collect $X_1(t)$ and $X_2(t)$:
\begin{align}
    \left(1-\sum_{i=1}^{n}a_iL^{i}\right)X_{1}(t)  + \left(-\sum_{i=1}^{n}b_iL^{i}\right)X_{2}(t) = \epsilon^{*}_{1}(t),
    \label{eq:ar_xy3}\\
    \left(-\sum_{i=1}^{n}c_iL^{i}\right)X_{1}(t) + \left(1-\sum_{i=1}^{n}d_iL^{i}\right)X_{2}(t) = \epsilon^{*}_{2}(t).
    \label{eq:ar_yx3}
\end{align}
We define the coefficients
$a(L) = 1-\sum_{i=1}^{n}a_iL^{i}$,
$b(L) = -\sum_{i=1}^{n}b_iL^{i}$,
$c(L) = -\sum_{i=1}^{n}c_iL^{i}$ and $d(L) = 1-\sum_{i=1}^{n}d_iL^{i}$,
and rewrite equations~\eqref{eq:ar_xy3} and~\eqref{eq:ar_yx3} into matrix form:
\begin{equation}
    \begin{pmatrix} 
    a(L) & b(L) \\ 
    c(L) & d(L) \\ 
    \end{pmatrix} \:
    \begin{pmatrix} 
    X_{1}(t) \\ 
    X_{2}(t) \\ 
    \end{pmatrix} =
        \begin{pmatrix} 
    \epsilon^{*}_{1}(t) \\ 
    \epsilon^{*}_{2}(t)
    \end{pmatrix}
    \label{eq:ar_matriz}
\end{equation}
where $a(0)=d(0)=1$ and $b(0)=c(0)=0$.

We apply the Fourier transform to equation~\eqref{eq:ar_matriz} in order to
switch to the frequency domain,
\begin{equation}
    \underbrace{
    \begin{pmatrix} 
    \Tilde{a}(\omega) & \Tilde{b}(\omega) \\ 
    \Tilde{c}(\omega) & \Tilde{d}(\omega) \\ 
    \end{pmatrix}}_{\pmb{A}(\omega)}
    \underbrace{
    \begin{pmatrix} 
    X_{1}(\omega) \\ 
    X_{2}(\omega) \\ 
    \end{pmatrix}}_{\mathbf{X}(\omega)} =
     \underbrace{
     \begin{pmatrix} 
    \epsilon_{1}^{*}(\omega) \\ 
    \epsilon_{2}^{*}(\omega)
    \end{pmatrix}}_{\pmb\Sigma(\omega)},
    \label{eq:ar_matriz2}
\end{equation}
where $\omega$ is the frequency and $\pmb{A}(\omega)$ is the coefficient matrix whose elements
are given by
\begin{align*}
\Tilde{a}(\omega) &= 1-\sum_{i=1}^{n}a_{i}\exp(-j\omega i),\\
\Tilde{b}(\omega) &= -\sum_{i=1}^{n}b_i\exp(-j\omega i),\\
\Tilde{c}(\omega) &= -\sum_{i=1}^{n}c_i\exp(-j\omega i),\\
\Tilde{d}(\omega) &= 1-\sum_{i=1}^{n}d_i\exp(-j\omega i).\\
\end{align*}

The expressions above are obtained by representing the lag operator in the spectral domain as $L^{i} = \exp{(-j\omega i)}$. This derives from the $z$-transform, where the representation of the $z$ variable\footnote{The lag operator $L$ is similar to the $z$-transform. However, $z$ is treated as a variable, and is often used in signal processing, while $L$ is an operator \cite{van2018signal}.} in the unit circle ($|z| = 1$) is $z^{-i} = \exp{(-j\omega i)}$ \cite{takalo2005tutorial,meddins2000introduction}.

To obtain the power spectra of $X_{1}(\omega)$ and $X_{2}(\omega)$, we first isolate
$\mathbf{X}(\omega)$ in equation~\eqref{eq:ar_matriz2}:
\begin{equation}
    \begin{pmatrix} 
    X_{1}(\omega) \\ 
    X_{2}(\omega) \\ 
    \end{pmatrix} =
    \underbrace{
    \begin{pmatrix} 
    H_{11}(\omega) & H_{12}(\omega) \\ 
    H_{21}(\omega) & H_{22}(\omega) \\ 
    \end{pmatrix}}_{\mathbf{H}(\omega)}
   \begin{pmatrix} 
    \epsilon^{*}_{1}(\omega) \\ 
    \epsilon^{*}_{2}(\omega)
    \end{pmatrix},
    \label{eq:ar_matriz3}
\end{equation}
where $\mathbf{H}(\omega) = \mathbf{A^{-1}}(\omega)$ is called the transfer matrix,
resulting in the following spectra:
\begin{equation}
\mathbf{S}(\omega) = \langle\mathbf{X}(\omega) \mathbf{X^{\dag}}(\omega)\rangle = \mathbf{H}(\omega)\pmb\Sigma(\omega) \mathbf{H^{\dag}}(\omega),
\label{eq:mat_spec}
\end{equation}
where $\langle . \rangle$ is the ensemble average, $\dag$ the transposed conjugate of the matrix,
and $\mathbf{S}(\omega)$ is the spectral matrix defined as:
\begin{equation}
    \mathbf{S}(\omega) = \begin{bmatrix} 
                    S_{11}(\omega) & S_{12}(\omega) \\
                    S_{21}(\omega)& S_{22}(\omega) 
                 \end{bmatrix}.
                 \label{eq:spec_mat_def}
\end{equation}

In equation~\eqref{eq:spec_mat_def}, $S_{11}(\omega)$ and $S_{22}(\omega)$ are called the autospectra, and
the elements $S_{12}(\omega)$ and $S_{21}(\omega)$ are called the cross-spectra.

We can expand the product in equation~\eqref{eq:mat_spec} to obtain $S_{11}(\omega)$ and $S_{22}(\omega)$
(see Section~B of the Appendix for details) as:
\begin{align}
    S_{11}(\omega) &= \underbrace{\Bar{H}_{11}(\omega)\Sigma_{11}\Bar{H}_{11}^{\dag}(\omega)}_\text{Intrinsic} \nonumber\\
    &+\underbrace{H_{12}(\omega)\left(\Sigma_{22}-\dfrac{\Sigma_{12}^2}{\Sigma_{11}^2}\right)H_{12}^{*}(\omega)}_\text{Causal},
    \label{eq:sxx}\\
    S_{ 22}(\omega) &= \underbrace{\hat{H}_{22}(\omega)\Sigma_{22} \hat{H}_{22}^{\dag}(\omega)}_\text{Intrinsic} \nonumber\\ &+\underbrace{\Bar{H}_{21}(\omega)\left(\Sigma_{11}-\dfrac{\Sigma_{21}^2}{\Sigma_{22}^2}\right)\Bar{H}_{21}^{*}(\omega)}_\text{Causal},
    \label{eq:syy}
\end{align}

\noindent where the symbols $\Bar{.}$ and $\hat{.}$ are used to differentiate the terms below
from the variables $H_{11}$, $H_{21}$, and $H_{22}$, as follows:
\begin{align*}
    \Bar{H}_{11}(\omega) &= H_{11}(\omega) + \Sigma_{12} H_{12}(\omega)\Sigma_{11},\\ \Bar{H}_{21}(\omega) &= H_{21}(\omega) + \Sigma_{12} H_{11}(\omega)\Sigma_{11},\\
    \hat{H}_{22}(\omega) &= H_{22}(\omega) + \dfrac{\Sigma_{12}}{\Sigma_{22}}H_{21}(\omega).
\end{align*}

Once we have the $S_{11}(\omega)$ and $S_{22}(\omega)$ spectra as the sum of an
\textit{intrinsic} and a \textit{causal} term, we may define indices to
quantify GC in frequency domain just as we did in the time domain (Section ~\ref{sec:gc_time}).
For instance, to calculate the causal index, we divide the spectra by their respective
intrinsic term in order to eliminate its influence.
Thus, the causality index  $I_{2 \rightarrow 1}(\omega)$ is defined as:
\begin{equation}
    I_{2 \rightarrow 1}(\omega) = \log\left(\dfrac{S_{11}(\omega)}{\Bar{H}_{11}(\omega)\Sigma_{11}\Bar{H}_{11}^{*}(\omega)}\right),
    \label{eq:ix2y}
\end{equation}
and analogously, $I_{1 \rightarrow 2}(\omega)$,
\begin{equation}
    I_{1 \rightarrow 2}(\omega) = \log\left(\dfrac{S_{22}(\omega)}{\hat{H}_{22}(\omega)\Sigma_{22} \hat{H}_{22}^{*}(\omega)}\right).
    \label{eq:iy2x}
\end{equation}   
The instantaneous causality index $I_{1 \leftrightarrow 2}(\omega)$ is defined as:
\begin{equation}
    I_{1 \leftrightarrow 2}(\omega) = \log\dfrac{\left(\Bar{H}_{11}(\omega)\Sigma_{11}\Bar{H}_{11}^{*}(\omega)\right)\left(\hat{H}_{22}(\omega)\Sigma_{22} \hat{H}_{22}^{*}(\omega)\right)}{\det(\mathbf{S}(\omega))}.
    \label{eq:ixy}
\end{equation}

In equations~\eqref{eq:ix2y} to~\eqref{eq:ixy}, we have one index for each value $\omega$ of
the frequency. Conversely, in the time domain there was a single index for
the GC between the two signals $X_1$ and $X_2$. Just as discussed in Section~\ref{sec:gc_time},
the indices $I_{2\rightarrow1}(\omega)$, $I_{1\rightarrow2}(\omega)$
and $I_{1\leftrightarrow2}(\omega)$ are greater than zero if there is any
relation between the time series. They are zero otherwise.

Just like in the time domain, the total GC in the frequency domain is the sum of its individual components:
\begin{align}
       I(\omega) &= I_{2 \rightarrow 1}(\omega) + I_{1 \rightarrow 2}(\omega) +   I_{1 \leftrightarrow 2}(\omega),\nonumber\\ &= \log\left(\dfrac{S_{11}(\omega)S_{22}(\omega)}{\det(\pmb S(\omega))}\right).
\label{eq:gc_total_spec}
\end{align}
The total GC is related to the so-called coherence $C_{12}(\omega)$
between signals (see Section~C of the Appendix):
\begin{equation}
    I(\omega) = -\log(1-C_{12}(\omega)).
    \label{eq:coh_totalGC}
\end{equation}

Moreover, we recover the GC in time domain through~\cite{ding200617, dhamala2008estimating}:
\begin{equation}
    F_{i \rightarrow j} = \dfrac{1}{\omega_{f}-\omega_{0}}
    \int_{\omega_{0}}^{\omega_{f}}I_{i \rightarrow j}(\omega)d\omega.
    \label{eq:freq_time}
\end{equation}

\section{Estimating Granger causality from data}

In the last two sections we have mathematically defined the GC in both time and frequency domains. Here, we discuss how to calculate GC.
In Section~\ref{sec:non_parametric}, we address a non-parametric estimation method that involves
computing the Fourier and wavelet transforms
of $X_{1}(t)$, and $X_{2}(t)$~\cite{daubechies1992ten,torrence1998,stoica2005spectral}.
In Section~D of the Appendix, we address the parametric estimation of GC, which involves fitting the signals
$X_{1}(t)$, and $X_{2}(t)$ to auto-regressive models (Section~\ref{sec::ar_models}).

\subsection{Calculating GC through Fourier and Wavelet Transforms}
\label{sec:non_parametric}

Here, we will give a numerical example for calculating and interpreting GC using a nonparametric
estimation approach based on Fourier and wavelet transforms~\cite{dhamala2008estimating}.
Our example consists of calculating the spectral matrix $\mathbf{S}(\omega)$
through the Fourier transform of the signals.
For two stationary\footnote{Stationarity, by definition, refers to time shift invariance of the underlying process statistics, which implies that all its statistical
moments are constant over time~\cite{woyczynski2019first}. There are several types of stationarity. Here, the required stationarity conditions for defining power spectral densities are constant means and that the covariance between any two variables apart by a given time lag is constant regardless of their absolute position in time. }
signals $X_1(t)$ and $X_2(t)$,
the $i,j$ element of the spectral matrix in equation~\eqref{eq:spec_mat_def} is

\begin{equation}
    S_{ij}(\omega) = \dfrac{\langle \Tilde{X}_i(\omega)\Tilde{X}_j^*(\omega)
    \rangle}{T},\:\:\:i = 1,2 \text{  and  } j = 1,2, 
    \label{eq:spec}
\end{equation}

\noindent where $T$ is the total duration of the signal, $\Tilde{X}_i(\omega)$ is the discrete Fourier transform of  $X_{j}(t)$  (calculated by a fast fourier transform, FFT, algorithm) and
 $\Tilde{X}_j^{*}(\omega)$ is its complex conjugate.

The variable
$\omega$ contains the values of the frequency in the interval $[0,f_{\rm max}]$
corresponding to where the FFT was calculated.
If $\Delta t$ is the sampling time interval of the original signals,
then the sampling frequency is $f_{\rm s}=1/\Delta t$ and $f_{\rm max}=f_{\rm s}/2$,
whereas the frequency interval contains $n_{\omega}=1+T/(2\Delta t)$ points.
Then, for $m$ signals ($m=2$ in our example), we have a total of $n_{\omega}$ spectral
matrices $\pmb S$ of dimensions $m\times m$. Recall that the diagonal
elements of $\pmb S$ are called the autospectra, whereas the other elements
are called the cross-spectra.

The transfer matrix, ${\pmb H}(\omega)$ and the covariance matrix $\pmb\Sigma$
are given by the decomposition of ${\pmb S}(\omega)$ into the product of
equation~\eqref{eq:mat_spec}. The Wilson algorithm~\cite{wilson1972, wilson1978,dhamala2008estimating} (see also Section~E of the Appendix) may be used for the decomposition of spectral matrices. 

After determining these two matrices, we may calculate the GC indices
through the direct application of equations~\eqref{eq:ix2y} to~\eqref{eq:ixy}.

For example, consider the autoregressive system studied in Ref.~\cite{dhamala2008estimating},
which is given by:
\begin{equation}
\begin{split}
    X_1(t) &= 0.55X_1(t-1) - 0.8X_1(t-2) + CX_2(t-1) + \epsilon_{1}(t),\\
    X_2(t) &= 0.55X_2(t-1) - 0.8X_2(t-2) + \epsilon_{2}(t).
\end{split}
\label{eq:ar1}
\end{equation}
Here, $X_1(t)$ and $X_2(t)$ are $AR(2)$. The variable $t$ is the time step index,
such that the actual time is $t'=t\,\Delta t=t/f_{\rm s}$. Besides, we know by construction that
$X_2(t)$ influences $X_1(t)$ through the coupling constant $C$
(although the opposite does not happen).
The terms $\epsilon_1(t)$ and $\epsilon_2(t)$ are defined to have variance
$\Sigma_{11} = \Sigma_{22} = 1$ and covariance $\Sigma_{12} = 0$ (they are
independent random processes).
To obtain a smooth power spectrum, we simulated $5000$ trials of the system in equation~\eqref{eq:ar1} and computed the average spectra across trials.
We set the parameters as $C=0.25$, $f_{\rm s} = 200$~Hz and $T=25$~s, resulting
in $5000$ data points. 

\begin{figure}[tp]
    \centering
    \includegraphics[scale=0.2]{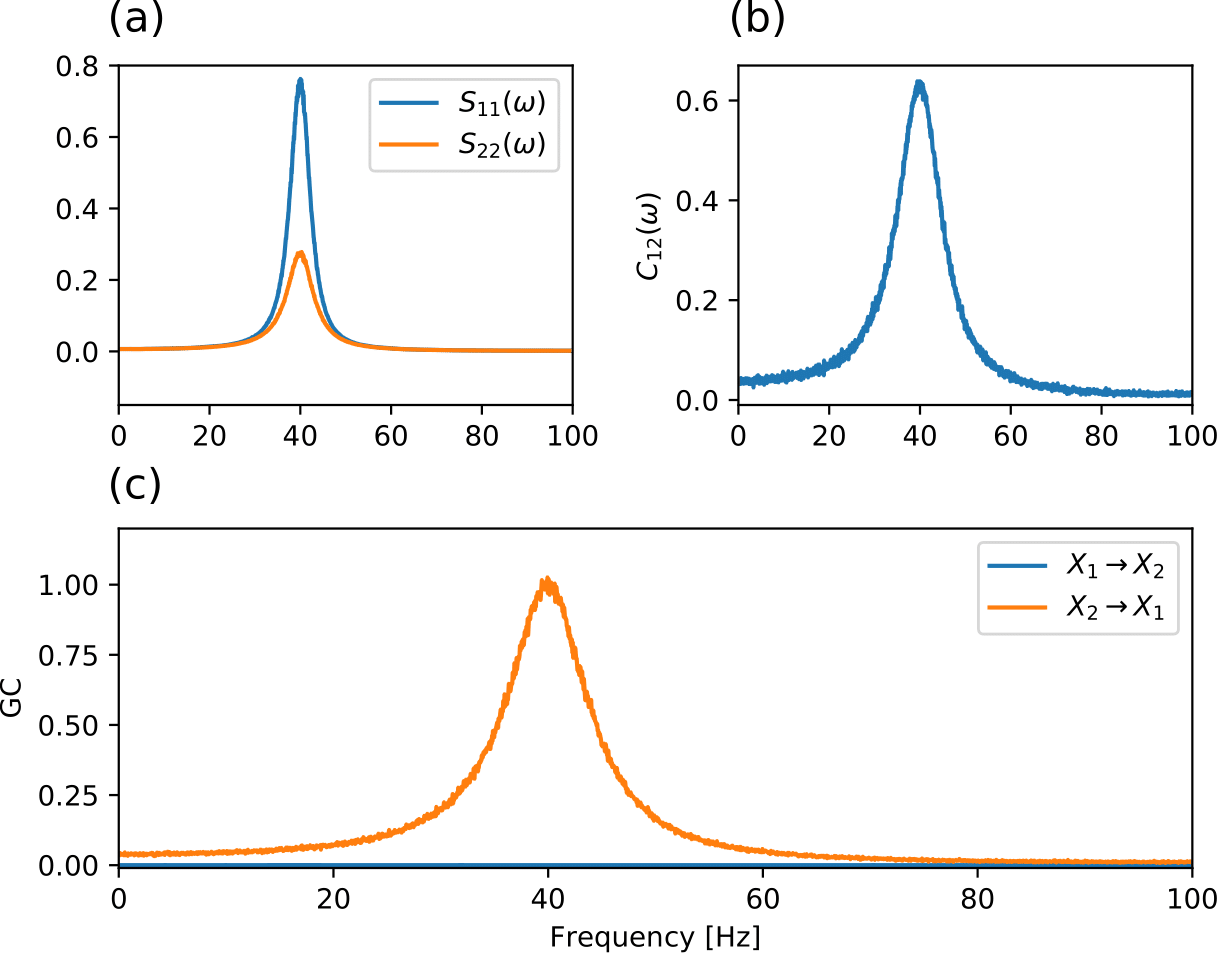}
    \caption{GC of a pair of autoregressive processes.
    GC for the system given in equation~\eqref{eq:ar1}: by construction, the process
    2 \textit{causes} 1 by providing it input through the coupling constant $C=0.25$.
    Parameters: total time $T=25$~s and sampling frequency $f_{\rm s} = 200$~Hz,
    resulting in $5000$ time steps.
    \textbf{a.} Spectral matrix components calculated via equation~\eqref{eq:spec}.
    \textbf{b.} Coherence between signals 1 and 2, equation.~\eqref{eq:coh_totalGC}.
    \textbf{c.} GC from 2 to 1 and 1 to 2, equation.~\eqref{eq:ix2y}
    and~\eqref{eq:iy2x}: a peak in 40~Hz in the $I_{2\rightarrow1}$ GC index
    indicates that 2 Granger-causes 1, whereas the flat zero $I_{1\rightarrow2}$
    shows, as expected, that 1 does not influence 2. The peak is in 40~Hz
    because process 2 has its main power in this frequency (see panel \textbf{a}).}
    \label{fig:Figura3}
\end{figure}

When $C=0$, $X_1(t)=X_2(t)$, both processes
are independent, and oscillate mainly in 40~Hz (Fig.~\ref{fig:Figura3}a).
For $C>0$, the process $X_1$ receives input from $X_2$, generating a
\textit{causal} coupling that is captured by the GC index in equations~\eqref{eq:ix2y}
and~\eqref{eq:iy2x}: a peak in 40~Hz in $I_{2\rightarrow1}$
indicates that process 2 (which oscillates in 40~Hz) is entering process 1
in this very frequency (Fig.~\ref{fig:Figura3}c). The flat spectrum
of $I_{1\rightarrow2}$ indicates that, on the other hand,
process 2 does not receive input from 1. The absolute value of $C$
changes the intensity of the GC peak.
The instant causality index, $I_{2\leftrightarrow1}(\omega)=0$ from
equation~\eqref{eq:ixy}, because $\Sigma_{12}=0$ for all $\omega$. The total
GC in the system is obtained from the spectral coherence, 
equation~\eqref{eq:coh_totalGC}. However, only the specific GC index reveal the
directionality of the inputs between 1 and 2.

This simple example illustrates
the meaning of causality in the GC framework: a Granger causal link is present if
a process runs under the temporal influence of the past of another signal.
We could have assumed $C$ as a time-varying function, $C(t)$, or even different
parameters for the autoregressive part of each process alone; or the
processes 1 and 2 could have been of different orders, implying in complex individual
power spectra. These scenarios are more usual for any real world
application~\cite{holme2012temporal,torrence1998}.
Then, instead of observing a clear peak
for the GC indices, we could observe a more complex pattern with peaks that
vary in time.

Instead of using the Fourier transform (which yields a single static spectrum for the
whole signal), we may use the Wavelet transform \cite{torrence1998, poularikas2010transforms,addison2017illustrated} to yield
time-varying spectra~\cite{bolzan2006transformada, domingues2016explorando}.
Then, the auto and cross-spectra from equation~\eqref{eq:spec}
may be written as
\begin{equation}
    S_{ij}(t, \omega) = \dfrac{\langle W_{i}(t, \omega)W^{*}_{j}(t, \omega) \rangle}{T},
    \label{eq:spec_wave}
\end{equation}
 where $W_{i}(t, \omega)$ is the Wavelet transform of $X_{i}(t)$ and
 $W_{i}^{*}(t, \omega)$ is its complex conjugate. To compute the Wavelet transform, we use a Morlet kernel \cite{torrence1998}, with scale $s=6$ oscillation cycles within a wavelet -- a typical value for this parameter~\cite{farge1992wavelet}. Similarly to what we did to the power spectrum, we measure the wavelet transforms for $5000$ trials of the system in equation~\eqref{eq:ar1} in order to average the results. It is important to stress that a wavelet transform is applicable in this case because ensemble averages are being taken. Otherwise, estimates would  be too unreliable for any meaningful inference.
 
 In practice, there is one matrix $\pmb S$
 for each pair $(t,\omega)$; or more intuitively, we have $n_T=T/\Delta t$ matrices
 $\pmb S(\omega)$, each one for a given time step $t$.
 The decomposition of $\pmb S(\omega)$
 in equation~\eqref{eq:mat_spec} is done through Wilson's
 algorithm. Then, we may calculate GC's indices via equations~\eqref{eq:ix2y} 
 to~\eqref{eq:ixy} for each of the $\pmb S(t,\omega)$ matrices with fixed $t$.
 This calculation results in $I_{2 \rightarrow 1}(\omega)$,
 $I_{1 \rightarrow 2}(\omega)$ and $I_{1 \leftrightarrow 2}(\omega)$
 for each time step $t$.
 Finally, we concatenate these spectra across the temporal dimension,
 yielding $I_{2 \rightarrow 1}(t,\omega)$, $I_{1 \rightarrow 2}(t,\omega)$ 
 and $I_{1 \leftrightarrow 2}(t,\omega)$.

For example, consider the same set of processes in equation~\eqref{eq:ar1},
but with time-varying $C(t)=0.25\,H(t_0-t)$, where $H(x)=1$ for $x\geq0$ (zero
otherwise) is the Heaviside step function. The parameter $t_0$
is the time step index in which the coupling from 2 to 1 is turned off.
This scenario is equivalent to having a set of concatenated constant $C$ processes,
such that the processes with $t>t_0$ have $C=0$. Then, we expect
the analysis in Fig.~\ref{fig:Figura3} to be valid for all the time steps
$t<t_0$, and no coupling should be detected whatsoever for $t>t_0$.

\begin{figure}[tp]
    \centering
    \includegraphics[scale=0.2]{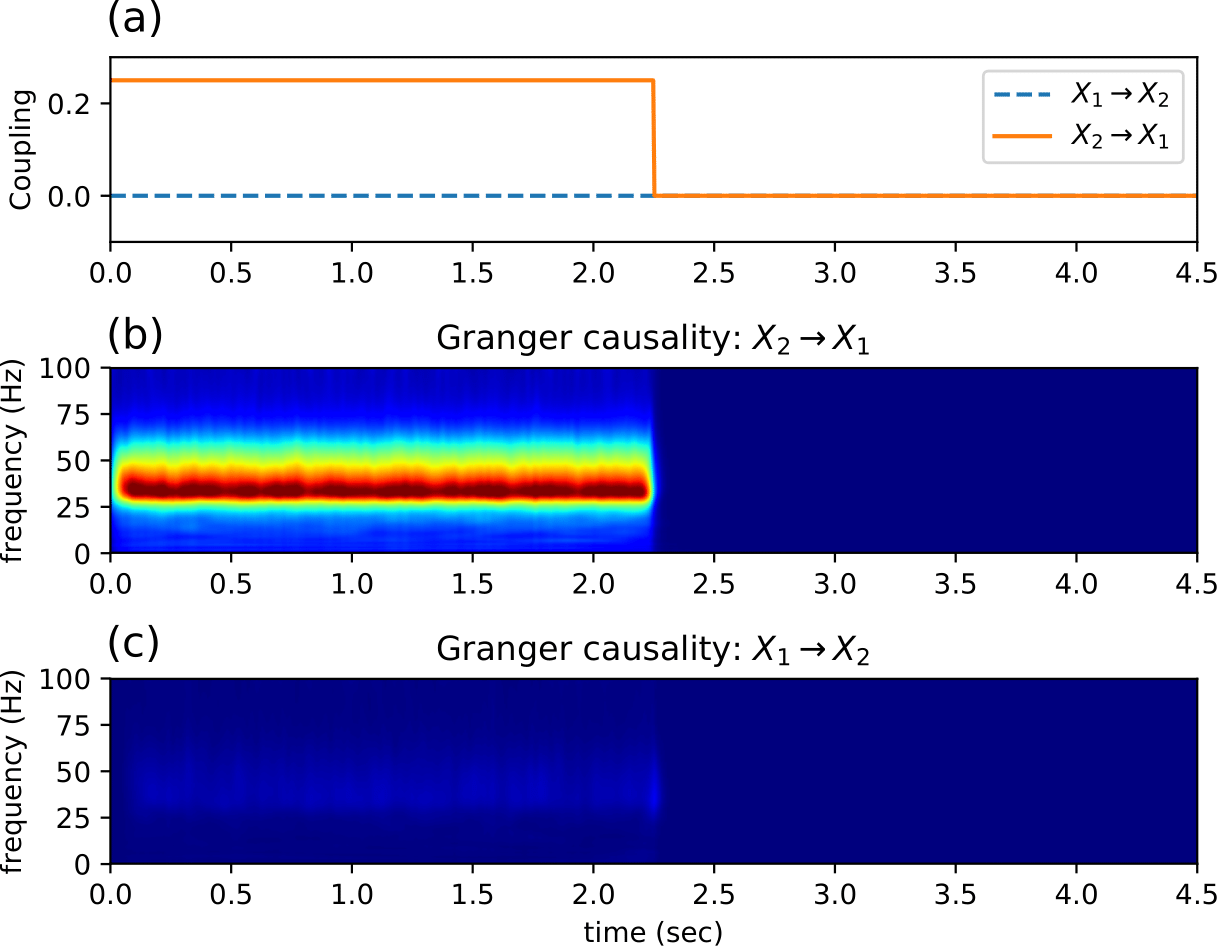}
    \caption{Time-varying GC in the frequency domain.
    GC of the system defined in equation~\eqref{eq:ar1}, but with
    time-varying $C(t)=0.25\,H(t_0-t)$. The spectral matrix is calculated
    via a Wavelet transform, equation~\eqref{eq:spec_wave}, and decomposed
    for each time step $t$, yielding a temporal decomposition
    of the frequencies of the signals.
    \textbf{a.} Coupling constant as function of time.
    \textbf{b.} GC index from 2 to 1, $I_{2 \rightarrow 1}(t,\omega)$.
    \textbf{c.} GC index from 1 to 2, $I_{1 \rightarrow 2}(t,\omega)$.}
    \label{fig:Figura4}
\end{figure}

That is exactly what is shown in Fig.~\ref{fig:Figura4}: a sharp transition
in the $I_{2 \rightarrow 1}(t,\omega)$ happens exactly at $t=t_0$
when $C$ is turned off. The index $I_{1 \rightarrow 2}(t,\omega)$ remains
zero for all the simulation. Again, this illustrates the meaning
of GC in our system: whenever there is a directional coupling
from a variable to another, there is nonzero GC in that link, in the example from signal $X_2(t)$ to $X_1(t)$.

\section{Conditional Granger Causality}\label{sec:cgc}

The concepts developed so far may be applied to a case with $m$ variables.
In this case, in order to try and infer the directionality\footnote{Whether $i$ Granger-causes $j$
or \textit{vice-versa}.}
of the interaction between two signals, in a system with $m$ signals, we may use the
so-called conditional Granger causality
(cGC)~\cite{geweke1984measures,chen2006freq,ding200617,malekpour2015cond}.
The idea is to infer the GC between signals $i$ and $j$ given the knowledge of
all the other $m-2$ signals of the system. This is done by comparing the variances
obtained considering only $i$ and $j$ to the variances obtained considering all the
other signals in the system.
The $AR$ model from Eqs~\eqref{eq:ar_xy} and~\eqref{eq:ar_yx}
ends up having a total of $m$ variables.

We may write the cGC in time domain as
\begin{equation}
    F_{i \rightarrow j|k,\dots,m},
\end{equation}
or in the frequency domain as
\begin{equation}
    I_{i \rightarrow j|k,\dots,m}(\omega).
\end{equation}
But one may ask:  ``isn't it simpler to just calculate the standard GC between every pair of
signals in the system, always reducing the problem to a two-variable case?''

\begin{figure}[tp]
     \centering
     \includegraphics[scale=0.3]{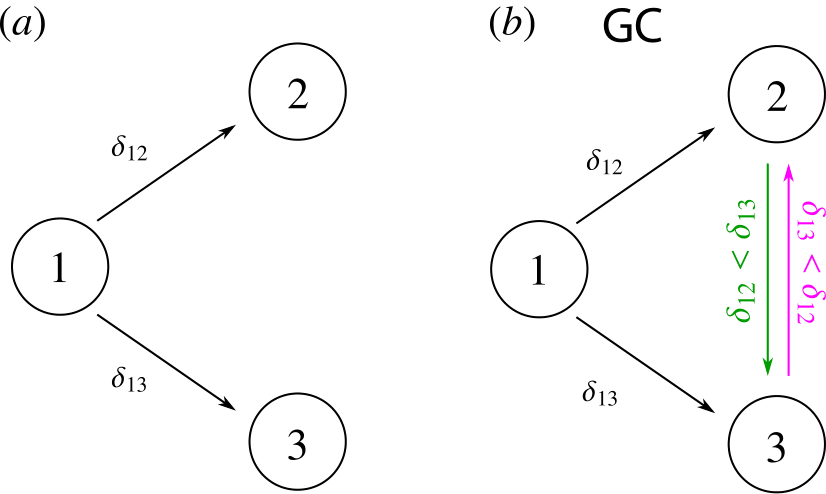}
     \caption{A system that GC fails to describe.
     \textbf{a.} Node 1 ($X_1(t)$) sends input to node 2 ($X_2(t)$) with delay $\delta_{12}$ and to node 3 ($X_3(t)$) with delay
     $\delta_{13}$.
     \textbf{b.} A simple GC calculation wrongly infer a link
     from $X_2(t)$ to $X_3(t)$ if $\delta_{12}<\delta_{13}$,
     or from $X_3(t)$ to $X_2(t)$ if $\delta_{13}<\delta_{12}$. These links are not physically present
     in the system and appear only due to the cross-correlation
     between $X_2(t)$ and $X_3(t)$ caused by the common input $X_1(t)$.}
     \label{fig:Figura5}
\end{figure}

To answer that question, consider the case depicted in Fig.~\ref{fig:Figura5}a:
node 1 ($X_1(t)$) sends input to node 2 ($X_2(t)$) with a delay $\delta_{12}$ and sends input to node 3 ($X_3(t)$)
with a delay $\delta_{13}$. Measuring the pairwise GC between $X_2(t)$ and $X_3(t)$ suggests
the existence of a coupling between them even if it does not physically exist
(as in Fig.~\ref{fig:Figura5}b).
This occurs because signals $X_2(t)$ and $X_3(t)$ are correlated due to their common input from $X_1(t)$,
and the simple pairwise GC between $X_2(t)$ and $X_3(t)$ fails to represent the correct relationship between the three nodes of Fig.~\ref{fig:Figura5}a.
The cGC solves this issue by considering the contribution of a third signal ($X_1(t)$ on this example) onto the analyzed pair ($X_2(t)$ and $X_3(t)$), as described below.

To describe the system in (Fig.~\ref{fig:Figura5}), equation~\eqref{eq:ar_matriz3} may be written as
\begin{equation}
    \begin{pmatrix} 
    X_{1}(\omega) \\ 
    X_{2}(\omega) \\ 
    X_{3}(\omega) \\
    \end{pmatrix} =
    \begin{pmatrix} 
    H_{11}(\omega) & H_{12}(\omega) & H_{13}(\omega) \\ 
    H_{21}(\omega) & H_{22}(\omega) & H_{23}(\omega)\\ 
    H_{31}(\omega) & H_{32}(\omega) & H_{33}(\omega)\\ 
    \end{pmatrix}
   \begin{pmatrix} 
    \epsilon^{*}_{1}(\omega) \\ 
    \epsilon^{*}_{2}(\omega) \\
    \epsilon^{*}_{3}(\omega)
    \end{pmatrix},
    \label{eq:ar_matriz4}
\end{equation}
where $X_{3}(t)$ has the noise term $\epsilon_{3}(t)$ with variance $\Sigma_{33}$.
The corresponding spectral matrix $\pmb S(\omega)$ is
\begin{equation}
    \pmb{S}(\omega) = \begin{bmatrix} 
                    S_{11}(\omega) & S_{12}(\omega) & S_{13}(\omega) \\
                    S_{21}(\omega)& S_{22}(\omega) & S_{23}(\omega) \\ 
                    S_{31}(\omega)& S_{32}(\omega) & S_{33}(\omega)
                 \end{bmatrix},
                 \label{eq:spec_mat_def3}
\end{equation}
and the noise covariance matrix is
\begin{equation}
\label{eq:covmat3}
    \pmb\Sigma = \begin{bmatrix} 
                    \Sigma_{11} & \Sigma_{12} & \Sigma_{13}  \\ 
                     \Sigma_{21} & \Sigma_{22} & \Sigma_{23} \\
                      \Sigma_{31} & \Sigma_{32} & \Sigma_{33}
                 \end{bmatrix}.
\end{equation}

We want to calculate the cGC from 
$X_2(t)$ to 
$X_3(t)$ given 
$X_1(t)$, \textit{i.e.}
$F_{2 \rightarrow 3|1}$ in the time domain and $I_{2 \rightarrow 3|1}(\omega)$ 
in the frequency domain. The first step is to build a partial system
from equation~\eqref{eq:ar_matriz4} ignoring the
coefficients related to the probe signal $X_2(t)$,
resulting in
the partial spectral matrix $\pmb S^{p}(\omega)$:
\begin{equation}
    \pmb{S}^{p}(\omega) = \begin{bmatrix} 
                    S_{11}(\omega) & S_{13}(\omega) \\
                    S_{31}(\omega)& S_{33}(\omega)
                 \end{bmatrix}.
                 \label{eq:part_spec_mat}
\end{equation}

From this partial system, we can calculate  $\pmb S^{p}(\omega)$ and $\pmb S(\omega)$ using the nonparametric
methods already discussed above. Suppose that for $\pmb S(\omega)$, we obtain 
the transfer matrix $\pmb H(\omega)$ and the covariance matrix $\pmb \Sigma$
(equation~\eqref{eq:covmat3}), whereas
for $\pmb S^{p}(\omega)$ we obtain the transfer matrix $\pmb G(\omega)$
and the covariance matrix $\pmb \rho$:

\begin{equation}
    \pmb{\rho} = \begin{bmatrix} 
                    \rho_{11} & \rho_{13} \\
                    \rho_{31}& \rho_{33} 
                 \end{bmatrix}.
\end{equation}

The matrices $\pmb H(\omega)$ and $\pmb \Sigma$
are $3\times3$. The matrices $\pmb G(\omega)$ and $\pmb \rho$ are always one dimension
less than the original ones, because they are built from the leftover rows and columns
of the original system without the coefficients of the probe signal.

In the time domain, $F_{2 \rightarrow 3|1}$ is defined as
\begin{equation}
    F_{2 \rightarrow 3|1} = \log\left(\dfrac{\rho_{33}}{\Sigma_{33}}\right),
    \label{eq:cgcgen}
\end{equation}
or, in general,
\begin{equation}
    F_{i \rightarrow j|k} = \log\left(\dfrac{\rho_{jj}}{\Sigma_{jj}}\right),
    \label{eq:cGCtindex}
\end{equation}
which is used to calculate the cGC from $i$ to $j$ given $k$, in time domain. Note 
that if the link between $i$ and $j$ is totally mediated by $k$, $\rho_{jj} = \Sigma_{jj}$, yielding 
$F_{i \rightarrow j|k}=0$. However, the standard GC between $i$ and $j$ would result in a link
between these variables. For our example in Fig.~\ref{fig:Figura5}, we
obtain $F_{2 \rightarrow 3|1}\gtrsim0$, meaning that the influence of $X_2(t)$ to
$X_3(t)$ is conditioned on signal $X_1(t)$, and hence is almost null.

In the frequency domain, we first must define the transfer matrix
$\pmb Q(\omega) = \pmb G(\omega)^{-1}\pmb H(\omega)$. However, the dimensions
of matrix $\pmb{G}(\omega)$ do not match the dimensions of matrix $\pmb{H}(\omega)$.
To fix that,
we add rows and columns from an identity matrix to the rows and columns
that were removed from the total system in equation~\eqref{eq:ar_matriz4}
when we built the partial system (\textit{i.e.}
we add the identity rows and columns to the rows and columns corresponding
to signal 
$X_2(t)$

that was

removed

for generating $\pmb S^{p}(\omega)$), such that:
\begin{equation}
    \mathbf{G}(\omega) = \begin{bmatrix} 
                    G_{11}(\omega) & G_{13}(\omega) \\
                    G_{31}(\omega)& G_{33}(\omega) 
                 \end{bmatrix} \Rightarrow
                 \begin{bmatrix} 
                    G_{11}(\omega) & 0 & G_{13}(\omega) \\
                    0 & 1 & 0 \\
                    G_{31}(\omega)& 0&G_{33}(\omega) 
                 \end{bmatrix}.
                 \label{eq:Gspec_mat_def}
\end{equation}

We can now safely calculate
$\pmb Q(\omega) = \pmb G(\omega)^{-1}\pmb H(\omega)$, from where we
obtain $I_{2 \rightarrow 3|1}(\omega)$:
\begin{equation}
    I_{2 \rightarrow 3|1}(\omega) = \log\left(\dfrac{\rho_{11}}{|Q_{11}(\omega)\Sigma_{11}Q_{11}^{\dag}(\omega)|}\right),
    \label{eq:cCG_spec}
\end{equation}

or, in general,
\begin{equation}
    I_{i \rightarrow j|k}(\omega) = \log\left(\dfrac{\rho_{jj}}{|Q_{jj}(\omega)\Sigma_{jj}Q_{jj}^{\dag}(\omega)|}\right).
    \label{eq:cGCfindex}
\end{equation}

\begin{figure}[tp]
     \centering
     \includegraphics[scale=.4]{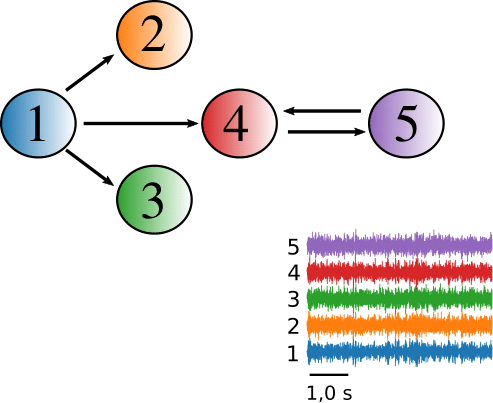}
     \caption{Many interacting components system. Illustration of a
     system with 5 interacting signals, having physical relations between them.
     We want to check whether GC or cGC is capable of capturing these interactions.}
     \label{fig:Figura6}
\end{figure}

To illustrate the procedures for determining cGC, consider the system defined in Fig.~\ref{fig:Figura6}
composed of 5 interacting elements~\cite{baccala2001partial}\footnote{Baccal\'{a} and Sameshima \cite{baccala2001partial} analysed this system using partial directed coherence and directed transfer function.}:
\begin{equation}
    \begin{split}
        X_{1}(t) &= 0.95\sqrt{2}X_{1}(t-1)-0.9025X_{1}(t-2) + \epsilon_{1}(t)\\
        X_{2}(t) &= 0.5X_{1}(t-2) + \epsilon_{2}(t)\\
        X_{3}(t) &= -0.4X_{1}(t-3) + \epsilon_{3}(t)\\
        X_{4}(t) &= -0.5X_{1}(t-2) + 0.25\sqrt{2}X_{4}(t-1) +\\
        & + 0.25\sqrt{2}X_{5}(t-1) + \epsilon_{4}(t)\\
        X_{5}(t) &= -0.25\sqrt{2}X_{4}(t-1) + 0.25\sqrt{2}X_{5}(t-1) + \epsilon_{5}(t).
    \end{split}
    \label{eq:baccala}
\end{equation}
Here, $X_{1}(t)$ sends its signal to $X_{2}(t)$, $X_{3}(t)$ and $X_{4}(t)$
with coupling intensities $0.5$, $-0.4$ and $-0.5$, respectively. Also, $X_{4}(t)$
sends input to $X_{5}(t)$ and \textit{vice-versa} with couplings
$-0.25\sqrt{2}$ and $0.25\sqrt{2}$ respectively. Note that $X_1$ sends signals
to $X_2$ and $X_4$ with 2 time steps of delay, and to $X_3$ with
3 time steps of delay. $X_4$ and $X_5$ exchange signals
with only 1 time step of delay.

\begin{figure}[tp]
     \centering
     \includegraphics[scale=0.3]{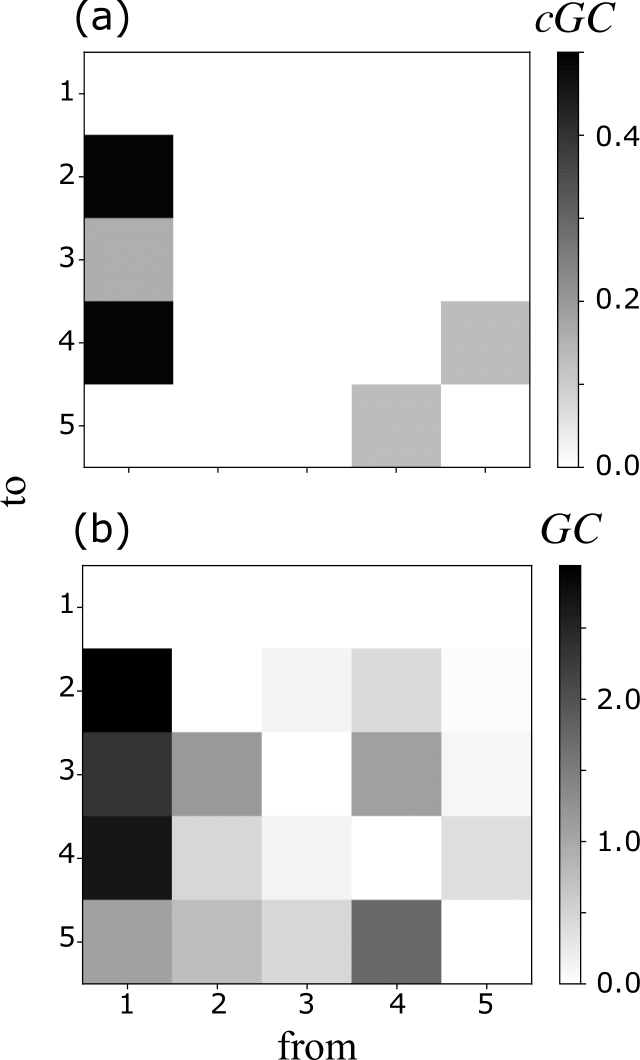}
     \caption{cGC and GC matrices in the time domain
     for a system of many interacting components.
     The system is given by equation~\eqref{eq:baccala}
     and is depicted in Fig.~\ref{fig:Figura6}.
     \textbf{a.} Conditional GC, equation~\eqref{eq:cGCtindex}, captures exactly the physical
     interactions of the system.
     \textbf{b.} Simple pairwise GC, equation~\eqref{eq:freq_time}, captures the interactions,
     but also captures underlying correlations coming from the hierarchy of the network.
     Real systems often do not have a clear cGC matrix as given in panel
     \textbf{a} due to second order effects. The covariance matrices
     $\pmb\Sigma$ and $\pmb\rho$ were calculated using the nonparametric method.
     }
     \label{fig:Figura7}
\end{figure}

\begin{figure*}[tp]
     \centering
     \includegraphics[scale=0.3]{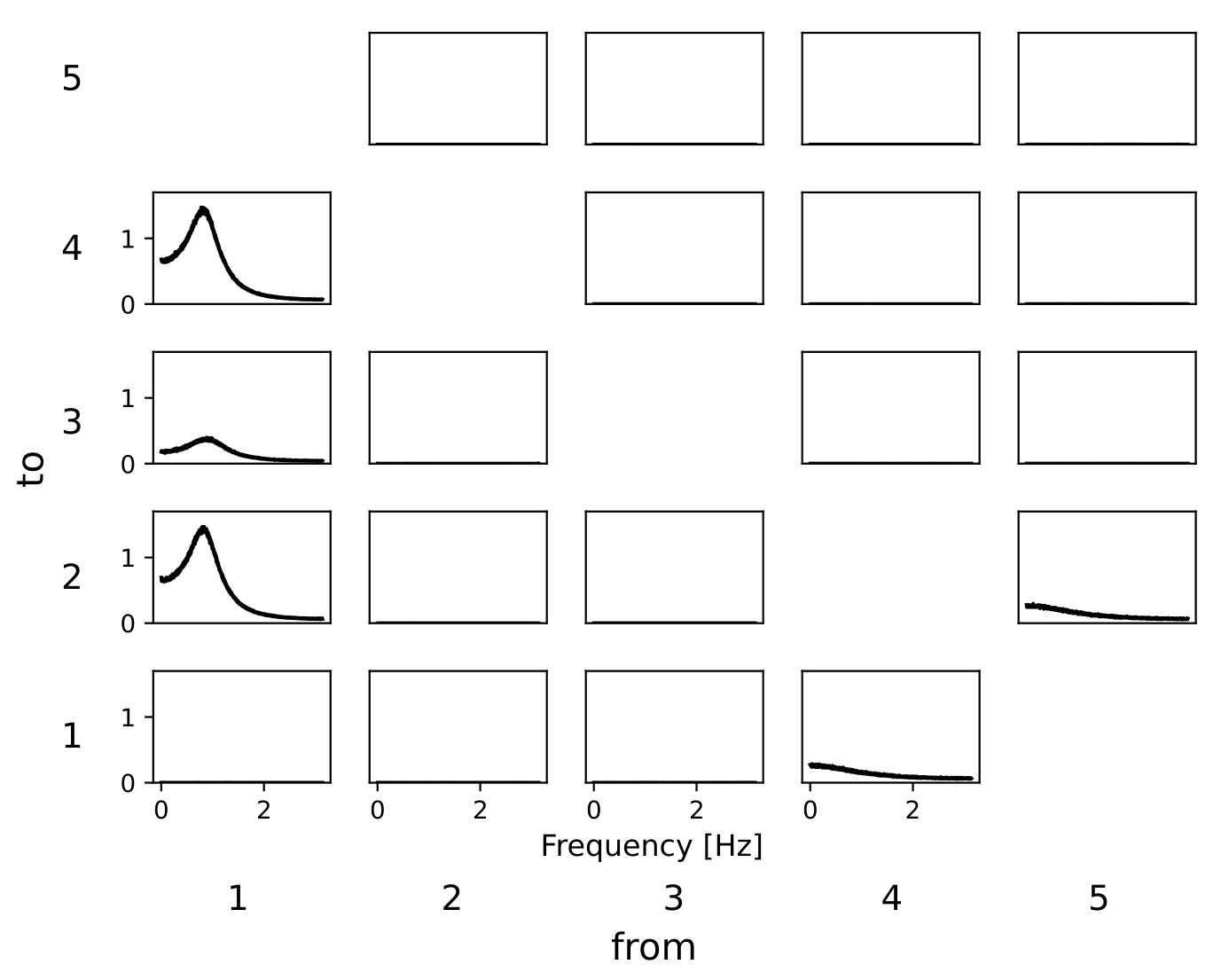}
     \caption{cGC and GC matrices in the frequency domain
     for a system of many interacting components.
     The same system as depicted Fig.~\ref{fig:Figura6}. Each panel
     in row $i$ and column $j$ corresponds to the cGC index between
     elements $i$ and $j$ in the frequency domain, equation~\eqref{eq:cGCfindex}.}
     \label{fig:Figura8}
\end{figure*}

Calculating the cGC index through equations~\eqref{eq:cGCtindex} and~\eqref{eq:cGCfindex},
we recover the expected structure of the network (Figs. \ref{fig:Figura7}a
and~\ref{fig:Figura8}, respectively).
The gray shades in Fig.~\ref{fig:Figura7} and the amplitude of the peaks in Fig.~\ref{fig:Figura8} are proportional to the coupling constants between each pair of elements.
For comparison, Fig.~\ref{fig:Figura7}b shows the simple pairwise GC,
which detects connections that are not physically present in the system. Again, this occurs
because the hierarchy of the network generates correlations between many pairs of
signals that are not directly connected, as discussed in the
example of Fig.~\ref{fig:Figura5}.

It is important to note that the cGC connectivity not always reflects the underlying
physical (or structural) connectivity between elements~\cite{seth2007granger}.
The example system in Fig.~\ref{fig:Figura6} is an illustrative simple case in which we obtained a neat result. However, real-world applications, such as
inferring neuronal connectivity from brain signals, result in a cGC matrix that is more noisy
due to multiple incoming signals and multiple delays. Thus, cGC is most generally
referred to as giving ``functional'' connectivity, instead of structural connectivity.

\section{Conclusion}
\label{sec:conc}

Granger causality is becoming increasingly popular as a method to determine
the dependence between signals in many areas of science. We presented its mathematical
formulation and showed examples of its applications in general systems of
interacting signals. This article also gives a contemporary scientific application
of the Fourier transform -- a subject that is studied in theoretical physics courses, but usually lacks practical applications in the classroom.
We also used wavelet transforms, which may motivate students to learn more about the
decomposition of signals in time and frequency domain, and its limitations through
the uncertainty principle.

We showed numerical examples, and explained them in an algorithmic way.
We included the inference of steady and time-varying coupling, and
the inference of connectivity in hierarchical networks via the conditional GC.
A limitation of the GC is that it is ultimately based on linear regression models
of stochastic processes (the AR models introduced in Section~\ref{sec::ar_models}).
Other measures, such as the transfer entropy, are more suitable to describe
nonlinear interactions, and do not need to be fitted to an underlying model.
Even though the nonparametric estimation of GC does not rely on
fitting, it is still a measure of linear interaction.
It is also possible to show that both GC and transfer entropy
yield the same results for Gaussian variables~\cite{barnett2009granger}.

In spite of the existing debate about what exactly the GC captures,
specially in the neuroscience community~\cite{stokes2017study,barnett2018solved}, 
GC has become a well-established measurement for the flux of information in
the nervous system~\cite{bressler2011wiener}. And here, we hope to have provided
the necessary tools to those who wish to learn the basic principles and applications
underlying GC.

\section{Supplemental Material}

All codes were developed in Python and are available in: \href{https://github.com/ViniciusLima94/pyGC}{https://github.com/ViniciusLima94/pyGC}.

\section{Acknowledgments}

This article was produced as part of the S. Paulo Research Foundation
(FAPESP) Research, Innovation and Dissemination Center
for Neuromathematics (CEPID NeuroMat, Grant No. 2013/07699-0). 
The authors also thank FAPESP support through Grants No. 2013/25667-8 (R.F.O.P.),
2015/50122-0 (A.C.R.),
2016/03855-5\\(N.L.K.), 
2017/07688-9 (R.O.S),
2018/20277-0 (A.C.R.)
and
2018/09150-9 (M.G.-S.).
V.L. is supported by a CAPES PhD scholarship.
A.C.R. thanks
financial support from the National Council of Scientific
and Technological Development (CNPq), Grant No. 306251/2014-0. This study was financed in part by the Coordenação de Aperfeiçoamento de Pessoal de Nível Superior - Brasil (CAPES) - Finance Code 001.



\bibliographystyle{unsrt}

\clearpage
\pagebreak

\appendix

\setcounter{page}{1}
\section*{Appendix}
\label{sec::apendice_yule}
\renewcommand{\thesubsection}{\Alph{subsection}}

\subsection{The Yule-Walker equations}
\renewcommand{\thefigure}{A\arabic{figure}}
\renewcommand{\theequation}{A\arabic{equation}}
\renewcommand{\thetable}{A\arabic{table}}
\setcounter{figure}{0}
\setcounter{equation}{0}
\setcounter{table}{0}

In order to derive the Yule-Walker set of equations, first we recall the definition of the  cross-correlation between two signals as given by equation~\eqref{eq:autocorrelation}:

\begin{equation}
\begin{split}
    R_{XY}(\tau) &= \frac{1}{N} \sum_{t=0}^{N+\tau-1}X(t)Y(t-\tau) \\&=\langle X(t)Y(t-\tau)\rangle,
    \end{split}
    \label{eq:autocorrelation}
\end{equation}

\noindent where $N$ is the number of data points, and $\tau$ the lag applied between the time series. In particular, if $X=Y$, $R_{XY}(\tau)$ is called the autocorrelation function, and dictates the dependency of future values of a time series with its past values \cite{golten1997understanding}, being maximal for $\tau = 0$.

Now let us consider the general AR process of order $n$:

\begin{equation}
    X(t) = \sum_{i=1}^{n}a_{i}X(t-i)+ \epsilon(t).
    \label{eq:ar_gen}
\end{equation}

Next, by multiplying equation~\eqref{eq:ar_gen} by $X(t-\tau)$, $\tau \geq 0$, taking the expected value of each term, and isolating the noise term in the right-hand side of this equation we end up with:

\begin{multline}
    \underbrace{\langle X(t)X(t-\tau)\rangle}_{R_{XX}(\tau)}
    -\sum_{i=1}^{n}a_{i}\underbrace{\langle X(t-i)X(t-\tau)\rangle}_{R_{XX}(\tau-i)} \\= \underbrace{\langle \epsilon(t)X(t-\tau)\rangle}_{R_{X\epsilon}(\tau)},
    \label{eq:ar_corr}
\end{multline}

\noindent where the terms indicated by the underbraces are precisely the autocorrelation of $X(t)$ ($R_{XX}$), and the cross-correlation between $X(t)$ and the noise $\epsilon(t)$ ($R_{X\epsilon}$).

The next step is to determine the term $R_{X\epsilon}(\tau-i)$ in equation~\eqref{eq:ar_corr}, to do so, first note that we can write $X(t-\tau)$ from equation~\eqref{eq:ar_gen} as:

\begin{equation}
X(t-\tau) = \sum_{i=1}^{n}a_{i}X(t-i-\tau) + \epsilon(t-\tau),
\end{equation}

\noindent which leads to:

\begin{equation}
\begin{split}
&R_{X\epsilon}(\tau) = \langle \epsilon(t)X(t-\tau)\rangle \\ &= \langle \epsilon(t)\sum_{i=1}^{n}a_{i}X(t-i-\tau) + \epsilon(t-\tau)\rangle \\&=
\langle \sum_{i=1}^{n}a_{i}\underbrace{X(t-i-\tau)\epsilon(t)}_{0} + \epsilon(t-\tau)\epsilon(t)\rangle \\&= \langle \epsilon(t-\tau)\epsilon(t) \rangle.
\end{split}
\label{eq:noise_corr}
\end{equation}

Note that we have used the fact that signal and the noise are always uncorrelated, hence $\sum_{i=1}^{n}X(t-i-\tau)\epsilon(t)$ is equal to zero.

It is known that the autocorrelation of a white noise process relates to its power spectrum via the Wiener-Khinchin theorem\footnote{The Wiener-Khinchin theorem relates the autocorrelation $R_{\epsilon\epsilon}(\tau)$ with its power spectrum $S(\omega)$ via the inverse Fourier transform: $R_{\epsilon\epsilon}(\tau) = \int_{-\infty}^{+\infty}\exp{(- i \omega \tau)}S(\omega)d\omega$ \cite{gardiner1985handbook}.}  \cite{laing2010stochastic}, and that $R_{X\epsilon}(\tau)$ is zero but for $\tau = 0$, where $R_{X\epsilon}(0) = \Sigma$, and $\Sigma$ is the noise variance. Said that, we can rewrite equation
~\eqref{eq:noise_corr} as:

\begin{equation}
  R_{X\epsilon}(\tau) = \langle \epsilon(t-\tau)\epsilon(t) \rangle = \begin{cases}
        0 \text{, if $\tau > 0$,}
        \\
        \Sigma \text{, if $\tau = 0$},
        \end{cases}
        \label{eq:noise_corr2}
 \end{equation}

\noindent and equation~\eqref{eq:ar_corr} as:

\begin{multline}
  R_{XX}(\tau) - \sum_{i=1}^{n}a_{i}R_{XX}(\tau -i) \\= \langle \epsilon(t-\tau)\epsilon(t) \rangle = \begin{cases}
        0 \text{, if $\tau > 0$,}
        \\
        \Sigma \text{, if $\tau = 0$}.
        \end{cases}
        \label{eq:yw1}
 \end{multline}
 
For $\tau > 0$, equation~\eqref{eq:yw1} can be written as:
 
 \begin{equation}
     R_{XX}(\tau) = \sum_{i=1}^{n}a_{i}R_{XX}(\tau -i).
     \label{eq:yw2}
 \end{equation}

\noindent Rewriting equation~\eqref{eq:yw2} as a matrix assigning
$\tau$ from 1 to $n$, we get:

\begin{multline}
    \underbrace{
    \begin{pmatrix} 
    R_{XX}(1)\\ 
    R_{XX}(2)\\
    \vdots\\
    R_{XX}(n)\\
    \end{pmatrix}}_{\pmb r}
    \\=
    \underbrace{
     \begin{pmatrix} 
    R_{XX}(0) & R_{XX}(1) & \dots & R_{XX}(n-1) \\ 
    R_{XX}(1) & R_{XX}(0) & \dots & R_{XX}(n-2) \\ 
    \vdots    & \vdots     & \ddots & \vdots     \\
    R_{XX}(n-1) & R_{XX}(n-2) & \dots & R_{XX}(0) \\ 
    \end{pmatrix}}_{\pmb R}
    \underbrace{
    \begin{pmatrix}
    a_{1}\\ 
    a_{2}\\
    \vdots\\
    a_{n}\\
    \end{pmatrix}}_{\pmb A}.
    \label{eq:ar_matrizYW}
\end{multline}

The set of $n$ equations and $n$ variables in equation~\eqref{eq:ar_matrizYW}, are known as the Yule-Walker equations. Each element in matrix $R$ is given by $R_{XX}(i-j)$, where $i = 0,\dots, n$, and $j = 0,\dots, n$ are the row, and column index, respectively. Note that $R$ is a symmetric matrix due to the fact that the autocorrelation function is itself symmetric, i.e., $R_{XX}(\tau) = R_{XX}(-\tau)$.

We can solve equation~\eqref{eq:ar_matrizYW} to find the coefficients matrix $\pmb A$ with:

\begin{equation}
    \pmb A = \pmb R^{-1} \pmb r,
    \label{eq:yw3}
\end{equation}

\noindent  where $\pmb R^{-1}$ is the inverse matrix of $\pmb R$. Once we find the coefficients using equation~\eqref{eq:yw3}, the noise variance $\Sigma$ can be estimated with equation~\eqref{eq:yw2} for $\tau = 0$.

In practice, if we rewrite equation~\eqref{eq:ar_gen} in the
following matrix form:
\vfill
\pagebreak
\begin{widetext}
\begin{equation}
    \underbrace{
    \begin{pmatrix} 
    X(n)\\ 
    X(n+1)\\
    \vdots\\
    X(N-1)\\
    \end{pmatrix}}_{\pmb x}
    =
    \underbrace{
     \begin{pmatrix} 
    X(n-1) & X(n-2) & \dots & X(0) \\ 
    X(n) & X(n-1) & \dots & X(1) \\  
    \vdots    & \vdots     & \ddots & \vdots     \\
    X(N-2) & X(N-3) & \dots & X(N-n) \\ 
        \end{pmatrix}}_{\pmb X}
   {\pmb A} + {\pmb \epsilon},
    \label{eq:data_matriz}
\end{equation}
\end{widetext}
\noindent where $\pmb\epsilon = [\epsilon(n), \dots, \epsilon(N-1)]^{T}$, is the noise column vector (superscript $T$ indicates matrix transpose). We can compute $\pmb r$ and $\pmb R$ in equation~\eqref{eq:ar_matrizYW} with:
\begin{equation}
    \pmb r = \frac{\pmb X^{T}\pmb x}{N},
    \label{eq:out}
\end{equation}
and,
\begin{equation}
    \pmb R = \frac{\pmb X^{T}\pmb X}{N}.
     \label{eq:in}
\end{equation}
Here, $\pmb x$ is a vector whose components are the
values of the process $X(t)$ at each time step $t$ starting at
$t=n$ (row 1) up to $t=N-1$ (row $N-n$);
and the matrix $\pmb X$ has each of its rows given by the
previous $n$ states of $X(t)$, i.e. from $X(t-1)$ in column 1 all the way
up to $X(t-n)$ in column $n$. Then, $\pmb A$ is a column vector of the coefficients
$a_i$ to be determined.

Note that, despite being possible to find the coefficients and the noise variance via the solution of the Yule-Walker equations, the idea of what order $n$ we should use to fit the $AR$ process is generally unknown. To estimate the order of the AR process usually we compute the Akaike information criterion \cite{akaike1998information, ding2016crude}, given by\footnote{The Akaike information criterion can be defined for more than one variable as well: $AIC(n) = 2\log(\det{(\pmb\Sigma)}) + 2nN_{\rm vars}^{2}/N$ where $\pmb \Sigma$ is the covariance matrix \cite{ding200617}.}
equation~\eqref{eq:akaike}:
\begin{equation}
    AIC(n) = N\log(\Sigma) + 2n,
    \label{eq:akaike}
\end{equation}
\noindent for different orders $n$, and use the order which minimizes $AIC(n)$ as the optimal one. Intuitively, the AIC can be thought as trying to balance the goodness of the fit measured through the term $\log(\Sigma)$ and the number of parameter used to fit the model, with $2n$ being a penalty term.

The procedure to estimate the correlation matrices $\pmb R$ and $\pmb r$  equation~\eqref{eq:ar_matrizYW} shown above was done for one variable, but it can be extended to any number of variables. For instance,
two variables obey the set of equations:
\begin{equation}
\begin{split}
    X_1(t) &= \sum_{i=1}^{n}a_{i,1}X_1(t-i) + \sum_{i=1}^{n}c_{i,2}X_2(t-i) + \epsilon_1(t),\\
    X_2(t) &= \sum_{i=1}^{n}a_{i,2}X_2(t-i) + \sum_{i=1}^{n}c_{i,1}X_1(t-i) + \epsilon_2(t).
    \label{eq:ar2_gen}
\end{split}
\end{equation}
where we may write
$\vec{a}_1$ as a column vector with components $a_{i,1}$ (the memory of the process 1
from $i$ time steps in the past),
and $\vec{c}_1$ as a column vector with components $c_{i,1}$ (the constant that couples
signal 1 from $i$ time steps in the past to signal 2 in the present).
Analogously,
we may define $\vec{a}_2$ and $\vec{c}_2$.
Using a procedure similar to the one before, we can also write these equations into
a matrix of the form:

\vfill\pagebreak
\begin{widetext}
\begin{equation}
    \underbrace{
    \begin{pmatrix} 
    X_{1}(n) & X_{2}(n)\\ 
    X_{1}(n+1) & X_{2}(n+1)\\
    \vdots & \vdots\\
    X_{1}(N-1) & X_{2}(N-1)\\
    \end{pmatrix}}_{\bar{\pmb x}}
    =
    \underbrace{
     \begin{pmatrix} 
     X_{1}(n-1) & \dots  & X_{1}(0)   & X_{2}(n-1) & \dots  & X_{2}(0) \\ 
     X_{1}(n)   & \dots  & X_{1}(1)   & X_{2}(n)   & \dots  & X_{2}(1) \\  
     \vdots     & \ddots & \vdots     & \vdots     & \ddots & \vdots   \\
     X_{1}(N-2) & \dots  & X_{1}(N-n) & X_{2}(N-2) & \dots  & X_{2}(N-n) \\
        \end{pmatrix}}_{\bar{\pmb X}}
   {\bar{\pmb A}} + {\bar{\pmb \epsilon}}.
    \label{eq:data_matriz2new}
\end{equation}
\end{widetext}

\noindent equation~\eqref{eq:data_matriz2new} is a
tensor-like version of equation~\eqref{eq:data_matriz}, where the quantities
are defined as
$\bar{\pmb x} = [{\pmb x}_1,{\pmb x}_2]$,
$\bar{\pmb X} = [{\pmb X}_1,{\pmb X}_2]$, and
$\bar{\pmb \epsilon} = [{\pmb \epsilon}_1,{\pmb \epsilon}_2]$,
such that $\bar{\pmb A}=[{\pmb A}_1,{\pmb A}_2]$
with\footnote{The $\vec{w}=\vec{u}\oplus\vec{v}$ operator defines the
direct sum of $\vec{u}$ and $\vec{v}$, such as if $\vec{u}$ and
$\vec{v}$ are column vectors, then
$\vec{w}=\bigl(\begin{smallmatrix}\vec{u}\\\vec{v}\end{smallmatrix}\bigr)$
is the vector made by stacking $\vec{u}$ on top of $\vec{v}$.}
${\pmb A}_1=\vec{a}_1\oplus\vec{c}_2$ and
${\pmb A}_2=\vec{c}_1\oplus\vec{a}_2$.
Notice that the matrix $\bar{\pmb A}$
now contains both their own memory
coefficients, $a_{i,1}$ and $a_{i,2}$, and the coupling coefficients,
$c_{i,1}$ and $c_{i,2}$, for any given lag in the past between
$i=1$ and $i=n$.
The noise vectors are
each of the form ${\pmb\epsilon}_1 = [\epsilon_1(n), \dots, \epsilon_1(N-1)]^{T}$
(the same for ${\pmb\epsilon}_2$).

With equation~\eqref{eq:data_matriz2new}, we can use equations~\eqref{eq:out} 
and~\eqref{eq:in} to compute the matrices $\bar{\pmb R}$ and $\bar{\pmb r}$, yielding the
Yule-Walker equations for the pair of processes:
\begin{widetext}
\begin{equation}
    \underbrace{
    \begin{pmatrix} 
    R_{11}(1) & R_{12}(1)\\
    \vdots & \vdots\\
    R_{11}(n) & R_{12}(n)\\
    R_{21}(1) & R_{22}(1)\\
    \vdots & \vdots\\
    R_{21}(n) & R_{22}(n)\\
    \end{pmatrix}}_{\bar{\pmb r}}
    =
   \underbrace{
     \begin{pmatrix} 
    R_{11}(0) & \dots & R_{11}(n-1) & R_{12}(0) & \dots & R_{12}(n-1) \\ 
    \vdots    & \ddots&\vdots       &\vdots     &\ddots &\vdots\\
    R_{11}(n-1)&\dots & R_{11}(0) & R_{12}(n-1) & \dots & R_{12}(0) \\ 
    R_{21}(0) & \dots & R_{21}(n-1) & R_{22}(0) & \dots & R_{22}(n-1) \\ 
    \vdots    & \ddots&\vdots       &\vdots     &\ddots &\vdots\\
    R_{21}(n-1)&\dots & R_{21}(0) & R_{22}(n-1) & \dots & R_{22}(0) \\ 
    \end{pmatrix}}_{\bar{\pmb R}}
   \bar{\pmb A}.
    \label{eq:YW2}
\end{equation}
\end{widetext}

Finally, we plug
these two matrices in equation~\eqref{eq:yw3} to find the coefficient matrix $\pmb A$ for
the process in equation~\eqref{eq:ar2_gen}.
Note that the matrices $\bar{\pmb r}$ and $\bar{\pmb R}$ are square matrices in which
the elements are either vectors (in $\bar{\pmb r}$) or other square
matrices (in $\bar{\pmb R}$):
\begin{align}
\bar{\pmb r} &=
\begin{pmatrix}
    {\pmb R}_{11}^{\tau} & {\pmb R}_{12}^{\tau}\\
    {\pmb R}_{21}^{\tau} & {\pmb R}_{22}^{\tau}\\
\end{pmatrix},
\label{eq:yweqtens1}\\
\bar{\pmb R} &=
     \begin{pmatrix} 
    {\pmb R}_{11}^{\tau,i} & {\pmb R}_{12}^{\tau,i} \\ 
    {\pmb R}_{21}^{\tau,i} & {\pmb R}_{22}^{\tau,i} \\ 
    \end{pmatrix},
    \label{eq:yweqtens2}
\end{align}
where the inner matrices of $\bar{\pmb r}$ are given by the column vectors
${\pmb R}_{jk}^{\tau} = [R_{jk}(\tau)]^T$ with elements ranging from
$\tau=1$ to $n$; and
the inner matrices of $\bar{\pmb R}$ are given by the square matrix
${\pmb R}_{jk}^{\tau,i} = [R_{jk}(\tau-i)]$ for $\tau=1$ to $n$
as the row index and $i=1$ to $n$ as the column index.
This system represents a set of equations
like equation~\eqref{eq:ar_matrizYW} for each combination of $j=1,2$ and $k=1,2$.

Hence, the 3-variables case would follow exactly the same procedure,
with
$\bar{\pmb x} = [{\pmb x}_1,{\pmb x}_2,{\pmb x}_3]$,
$\bar{\pmb X} = [{\pmb X}_1,{\pmb X}_2,{\pmb X}_3]$, and
$\bar{\pmb \epsilon} = [{\pmb \epsilon}_1,{\pmb \epsilon}_2,{\pmb \epsilon}_3]$,
such that $\bar{\pmb A}=[{\pmb A}_1,{\pmb A}_2,{\pmb A}_3]$ with
${\pmb A}_1=\vec{a}_1\oplus\vec{c}_2\oplus\vec{c}_3$,
${\pmb A}_2=\vec{c}_1\oplus\vec{a}_2\oplus\vec{c}_3$, and
${\pmb A}_3=\vec{c}_1\oplus\vec{c}_2\oplus\vec{a}_3$.
These settings generate a Yule-Walker equation for a $3\times3$
matrix of matrices, similarly to equations~\eqref{eq:yweqtens1}
and~\eqref{eq:yweqtens2}. And so on and so forth for
the case of $N_{\rm vars}$ variables. It is easy to see
that the dimensions of the matrices involved in the Yule-Walker equations
grow exponentially fast with the number of variables involved.
Finally, for the multi-variables case, the coefficient matrix $\pmb A$ can be obtained in the same way as for the single variable case, using equation~\eqref{eq:yw3}.

\subsection{Derivation of the autospectra}
\renewcommand{\thefigure}{B\arabic{figure}}
\renewcommand{\theequation}{B\arabic{equation}}
\renewcommand{\thetable}{B\arabic{table}}
\setcounter{figure}{0}
\setcounter{equation}{0}
\setcounter{table}{0}

In order to obtain the expressions for autospectra and cross-spectra given in equations~(22)-(23)
following the method introduced by Geweke~\cite{geweke1982measurement}, we start by rewriting the product of the matrices given in equation~(20) as:
\begin{equation}
\begin{split}
    &\begin{bmatrix}
        S_{11}(\omega) & S_{12}(\omega) \\ 
        S_{21}(\omega) & S_{22}(\omega) \\ 
    \end{bmatrix} =\\
    &\begin{bmatrix} 
        H_{11}(\omega) & H_{12}(\omega) \\ 
        H_{21}(\omega) & H_{22}(\omega) \\ 
    \end{bmatrix}
    \begin{bmatrix}
        \Sigma_{11} & \Sigma_{12} \\
        \Sigma_{21} & \Sigma_{22} 
    \end{bmatrix}
    \begin{bmatrix} 
        H_{11}^{*}(\omega) & H_{21}^{*}(\omega) \\ 
        H_{12}^{*}(\omega) & H_{22}^{*}(\omega) \\ 
    \end{bmatrix}.
    \end{split}
    \label{eq:AP_Spec}
\end{equation}

By solving equation~\eqref{eq:AP_Spec}, we obtain the autospectra for $X_1(t)$ as:
\begin{equation}
\begin{split}
    S_{11}(\omega) & = H_{11}(\omega) \Sigma_{11} H_{11}^{*}(\omega) + H_{12}(\omega) \Sigma_{21} H_{11}^{*}(\omega) + \\
    & + H_{11}(\omega) \Sigma_{12} H_{12}^{*}(\omega) + H_{12}(\omega) \Sigma_{22} H_{12}^{*}(\omega).
\end{split}
\label{eq:AP_Spec_X1}
\end{equation}
Since $\Sigma_{12} = \Sigma_{21}$, equation~\eqref{eq:AP_Spec_X1} can be rewritten as:
\begin{equation}
\begin{split}
    S_{11}(\omega) & = H_{11}(\omega) \Sigma_{11} H_{11}^{*}(\omega)
    + H_{12}(\omega) \Sigma_{22} H_{12}^{*}(\omega) \\
    & + \Sigma_{12} (H_{12}(\omega) H_{11}^{*}(\omega) + H_{11}(\omega) H_{12}^{*}(\omega)).
\end{split}
\label{eq:AP_Spec_X1_simplified1}
\end{equation}
Note that $(H_{12}(\omega) H_{11}^{*}(\omega) + H_{11}(\omega) H_{12}^{*}(\omega)) = \\2 {\rm Re} \left(H_{11}(\omega) H_{12}^{*}(\omega)\right)$, and therefore, we can rewrite the autospectra for $X_{1}(t)$ as:
\begin{equation}
\begin{split}
    S_{11}(\omega) & = H_{11}(\omega) \Sigma_{11} H_{11}^{*}(\omega)
    + H_{12}(\omega) \Sigma_{22} H_{12}^{*}(\omega)\\
    & + 2 \Sigma_{12} {\rm Re} \left(H_{11}(\omega) H_{12}^{*}(\omega)\right).
\end{split}
\label{eq:s11}
\end{equation}

In the case where $\Sigma_{12} = 0$ in equation~\eqref{eq:s11}, only two terms will remain; the first can be seen as an intrinsic term involving only the variance of $\epsilon_{1}^{*}(t)$ and the terms of the transfer matrix related only to $X_{1}(t)$, and the second as a causal term involving only the variance of $\epsilon_{2}^{*}(t)$ and the terms of the transfer matrix regarding the relation between $X_{1}(t)$ and $X_{2}(t)$. 

This separation between intrinsic and causal terms is the key for defining the GC index in frequency domain described by equations~(24)--(26).

On the other hand, if $\Sigma_{12} > 0$, we will have a third term resulting from the influence that correlated noise exert on the spectra.
To overcome this problem, we consider the transformation introduced by Geweke~\cite{geweke1982measurement} which makes possible to remove the cross term\\ $2 \Sigma_{12} {\rm Re} \left(H_{11} H_{12}^{*}\right)$ in equation~\eqref{eq:s11}. The procedure consists in multiplying both sides of equation~(18) by the following matrix $\pmb P$:
\begin{equation}
\pmb P = 
    \begin{bmatrix} 
        1 & 0 \\ 
        -\Sigma_{22}/\Sigma_{11} & 1 \\ 
    \end{bmatrix}   , 
\label{eq:MP}
\end{equation}
which gives the following equation:
\begin{equation}
    \underbrace{
    \begin{pmatrix} 
    a(\omega) & b(\omega) \\ 
    \Tilde{c}(\omega) & \Tilde{d}(\omega) \\ 
    \end{pmatrix}}_{\pmb{\Tilde{A}}(\omega)}
    \begin{pmatrix}
    X_{1}(\omega) \\ 
    X_{2}(\omega) \\ 
    \end{pmatrix} =
     \begin{pmatrix} 
    \epsilon_{1}^{*}(\omega) \\ 
    \Tilde{\epsilon}_{2}^{*}(\omega)
    \end{pmatrix},
    \label{eq:ar_matriz5}
\end{equation}
where $\Tilde{c}(\omega) = c(\omega) - \dfrac{\Sigma_{22}}{\Sigma_{11}} a(\omega)$, $\Tilde{d}(\omega) = d(\omega) - \dfrac{\Sigma_{22}}{\Sigma_{11}} b(\omega)$ e $\Tilde{\epsilon}_{2}^{*}(\omega) = \epsilon_{2}^{*}(\omega) - \dfrac{\Sigma_{22} }{\Sigma_{11}} \epsilon_{1}^{*}(\omega)$.
The newly obtained tranfer matrix $\Tilde{\pmb H}(\omega)$ resulting from this transformation will be therefore the inverse of the coefficient matrix $\pmb{\Tilde{A}}(\omega)$.

For example, the inverse matrix of $\pmb A(\omega)$ with dimensions $2\times2$ is:
\begin{equation}
    \pmb H(\omega) = \pmb A(\omega)^{-1} = \dfrac{1}{\det\left( \pmb A \right)}
    \begin{pmatrix} 
    d(\omega) & -c(\omega) \\ 
    -b(\omega) & a(\omega) \\ 
    \end{pmatrix}.
    \label{eq:invA}
\end{equation}
Therefore, the elements of the transfer matrix $\Tilde{\pmb H}(\omega)$, given the fact that $\det\left(\pmb A(\omega)\right) = \det\left(\pmb{\Tilde{A}}(\omega)\right)$ are:
\begin{equation}
\begin{split}
    & \Tilde{H}_{11} = \dfrac{1}{\det\left(\pmb{A}\right)} \Tilde{d} = \dfrac{d(\omega)}{\det\left(\pmb{A}\right)} - \dfrac{\Sigma_{22}}{\Sigma_{11}} \dfrac{b(\omega)}{\det\left(\pmb{A}\right)} = H_{11} + \dfrac{\Sigma_{22}}{\Sigma_{11}} H_{12}, \\
    & \Tilde{H}_{12} = H_{12},\\
    & \Tilde{H}_{21} = \dfrac{1}{\det\left(\pmb{A}\right)} \Tilde{c} = \dfrac{c(\omega)}{\det\left(\pmb{A}\right)} - \dfrac{\Sigma_{22}}{\Sigma_{11}} \dfrac{a(\omega)}{\det\left(\pmb{A}\right)} = H_{21} + \dfrac{\Sigma_{22}}{\Sigma_{11}} H_{22}, \\
    & \Tilde{H}_{22} = H_{22}.\\
\end{split}
\end{equation}

Due to the fact that $\text{cov}(\epsilon^{*}_1,\Tilde{\epsilon}^{*}_2) = 0$ and $\text{Var}(\Tilde{\epsilon}^{*}_2) = \Sigma_{22} -\Sigma_{12}^2/\Sigma_{11}$, and considering the spectral matrix in terms of $\Tilde{\pmb H}(\omega)$ such that $\pmb S(\omega) = \Tilde{\pmb H}(\omega)
\pmb\Sigma \Tilde{\pmb H}^{\dag}(\omega)$ we have:
\begin{equation}
    S_{11}(\omega) = \Tilde{H}_{11}(\omega) \Sigma_{11} \Tilde{H^{*}}_{11}(\omega) + \Tilde{H}_{12}(\omega) \Sigma_{22} \Tilde{H}^{*}_{12}(\omega),
\end{equation}
as the sum of the intrinsic and causal terms.

Similarly, to obtain the autospectra for $S_{22}(\omega)$ we follow the same procedure considering the tranpose of the transformation matrix $\mathbf{P}$.

\subsection{Relationship between spectral coherence and Granger causality}
\label{sec::apendice_b}
\renewcommand{\thefigure}{C\arabic{figure}}
\renewcommand{\theequation}{C\arabic{equation}}
\renewcommand{\thetable}{C\arabic{table}}
\setcounter{figure}{0}
\setcounter{equation}{0}
\setcounter{table}{0}

Based on the auto-spectra and cross-spectra defined in equation~(30), we can define a spectral correlation index
called coherence ($C_{ij}(\omega)$) given by equation~\eqref{eq:coh}.

\begin{equation}
    C_{ij}(\omega) = \dfrac{|S_{ij}(\omega)|^{2}}{S_{ii}(\omega)S_{jj}(\omega)},
    \label{eq:coh}
\end{equation}
where $C_{ij}(\omega)$ is limited in $0 \leq C_{ij}(\omega) \leq 1$ \cite{shumway2005timeseries,kramer2013introduction}.

Furthermore, it is possible to establish a relationship between coherence and total causality given by equation~(27). Here we show how to obtain the total causality as a function of $C_{ij}(\omega)$, where the determinant of the matrix $\pmb S(\omega)$

is given by:
\begin{equation}
    \det(\pmb S(\omega)) = S_{11}(\omega) S_{22}(\omega) - S_{12}(\omega) S_{21}(\omega).
    \label{eq:AP_detS}
\end{equation}

Due to the fact that $S_{21}(\omega) = S_{12}^{*}(\omega)$, and that  $|S_{12}(\omega)|^2 = S_{12}(\omega) S_{12}(\omega)^{*}$, equation~\eqref{eq:AP_detS} can be rewritten as:

\begin{align}
    \det(\pmb S(\omega)) 
    & = S_{11}(\omega) S_{22}(\omega) - S_{12}(\omega) S_{12}^*(\omega),   \nonumber   \\
    & = S_{11}(\omega) S_{22}(\omega) - |S_{12}|^2.
\label{eq:AP_detS2}
\end{align}

Lastly, by inverting the arguments in the logarithm of equation~(27) and applying equation~\eqref{eq:AP_detS2}, the total Granger causality in frequency domain ($I(\omega)$) can be written as a function of spectral coherence ($C_{ij}(\omega)$) as:
\begin{align}
    I(\omega) 
    & = \log \left( \left( \dfrac{\det(S(\omega))}{S_{11}(\omega) S_{22}(\omega)}\right)^{-1} \right),   \nonumber    \\
    & = - \log \left( \dfrac{S_{11}(\omega) S_{22}(\omega) - |S_{12}|^2}{S_{11}(\omega) S_{22}(\omega)}\right),     \nonumber  \\
    & = - \log \left( 1 - \underbrace{\dfrac{|S_{12}|^2}{S_{11}(\omega) S_{22}(\omega)}}_{C_{12}(\omega)}\right).
    \label{eq:AP_Itot2}
\end{align}

\subsection{Parametric estimation of the GC}\label{sec:par_gc}
\renewcommand{\thefigure}{D\arabic{figure}}
\renewcommand{\theequation}{D\arabic{equation}}
\renewcommand{\thetable}{D\arabic{table}}
\setcounter{figure}{0}
\setcounter{equation}{0}
\setcounter{table}{0}

To illustrate the procedure for estimating the GC parametrically, consider the same example used in Section~5.1, and given by equation~(31).

In order to compute the spectral GC between $X_{1}(t)$, and $X_{2}(t)$ parametrically, we first have to estimate the $AR$ coefficients by solving the Yule-Walker equations (Section~A of the Appendix). Assuming that we know the model order to be $n=2$, equation~\eqref{eq:data_matriz2new} can be written as:

\begin{equation}
\footnotesize
    \begin{split}
    &\underbrace{
    \begin{pmatrix} 
    X_{1}(2) & X_{2}(2)\\ 
    X_{1}(3) & X_{2}(3)\\
    \vdots & \vdots\\
    X_{1}(N-1) & X_{2}(N-1)\\
    \end{pmatrix}}_{\bar{\pmb x}}
    =\\&=
    \underbrace{
     \begin{pmatrix} 
    X_{1}(1) & X_{1}(0) & X_{2}(1) & X_{2}(0) \\ 
    X_{1}(2) & X_{1}(1) & X_{2}(2) & X_{2}(1) \\  
    \vdots    & \vdots     & \vdots & \vdots     \\
    X_{1}(N-2) & X_{1}(N-3) & X_{2}(N-2) & X_{2}(N-3) \\
        \end{pmatrix}}_{\bar{\pmb X}}
   \bar{\pmb A} + \bar{\pmb \epsilon}.
    \label{eq:data_matriz2}
    \end{split}
\end{equation}

Next, we can compute the coefficients of the AR series using equations~\eqref{eq:yw3},\eqref{eq:out}-\eqref{eq:in}.

In Table~\ref{Tab::fit_ar2} we show the coefficients, and the covariance matrix obtained by solving the Yule-Walker equations for $n=2$. The values displayed are an average for $5000$ trials of the system given by equation~(31). We use such a large number of trials
for consistency with the method employed to obtain Fig.~\ref{fig:Figura3},
however this method should converge in less trials than the non-parametric
one.

\begin{table}[!bp]
	\begin{center}
		\begin{tabular}{ccccc}
			
			\hline 
		    &$X_{1}(t-1)$ & $X_{1}(t-2)$ & $X_{2}(t-1)$ & $X_{2}(t-2)$\\
		    \hline 
		    $X_{1}(t)$ & $0.55$ & $-0.80$ & $0.25$ & $0$\\
		    $X_{2}(t)$ & $0$ & $0$ & $0.55$ &$-0.80$\\
		    \hline
		    \\
		    \multicolumn{5}{c}{$ \pmb\Sigma = 
		     \centering
		    \begin{bmatrix} 
                    0.999 & 0 \\
                    0 & 0.999 
                 \end{bmatrix}.$}\\
                 \\
		    \hline 
			
		\end{tabular}
		\caption{\label{Tab::fit_ar2}Coefficient matrix $\pmb A^{T}$, and noise covariance matrix $\pmb\Sigma$, estimated through solving the Yule-Walker equations for the system given by equation~(31). The parameters were averaged for $5000$ trials.
		}
	\end{center}
\end{table}

With the coefficient matrix $\pmb A$, the transfer matrix $\pmb H(\omega)$ can be obtained with equation~\eqref{eq:transfer_matrix_coef}.

\begin{equation}
    \pmb H(\omega) = \left(\pmb I - \sum_{\tau=1}^{n=2} \pmb A(\tau)e^{\frac{-j\omega \tau}{f_{s}}} \right)^{-1},
    \label{eq:transfer_matrix_coef}
\end{equation}

\noindent where, $\pmb I$ is the identity matrix with dimensions $N_{\rm vars}$, $\pmb A(\tau)$ is the coefficient matrix for lag $\tau$, for our example:

\begin{equation}
    \pmb A(\tau=1) = 
		     \centering
		    \begin{bmatrix} 
                    0.55 & 0.25 \\
                    0 & 0.55
                 \end{bmatrix},
\end{equation}

\noindent and,
 
\begin{equation}
    \pmb A(\tau=2) = 
		     \centering
		    \begin{bmatrix} 
                    -0.80 & 0 \\
                    0 & -0.80
                 \end{bmatrix},
\end{equation}

\noindent $j$ the imaginary unit, and $\omega$ the frequencies of the signal (computed as explained in Section~5.1).

The spectral matrix $\pmb S(\omega)$, can be obtained using $\pmb H(\omega)$, and $\pmb \Sigma$ via equation~(20). Subsequently, the GC components can be obtained using equations~(24)-(26).

\begin{figure}[!bp]
    \centering
    \includegraphics[scale=0.22]{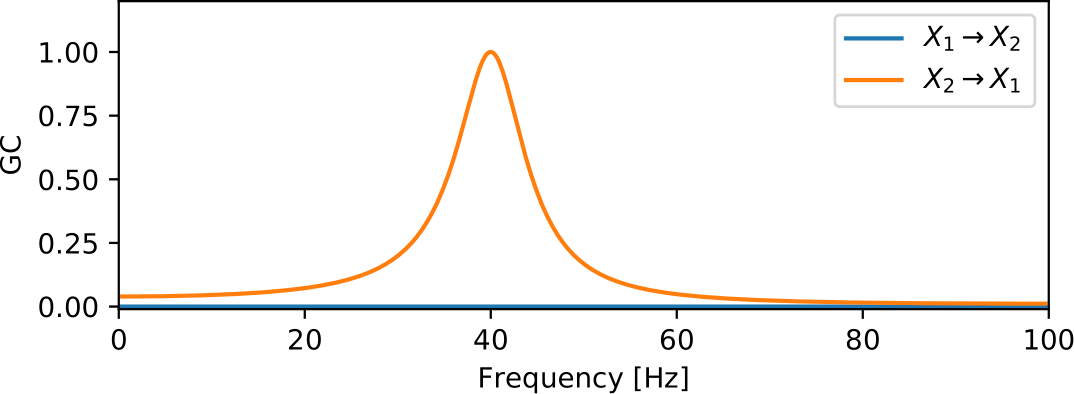}
    \caption{GC estimated parametrically for the system in equation~(31). The result is in accordance to the one obtained in Fig.~3(c), where GC was obtained with non-parametric estimation, and captures the directionality of the information flow imposed to the model. }
    \label{fig:Figura9}
\end{figure}

In Fig.~\ref{fig:Figura9}, we show the GC estimated in the frequency domain from $X_{1}\rightarrow X_{2}$, and $X_{2}\rightarrow X_{1}$.

As can be seen in Fig.~\ref{fig:Figura9}, the GC in fact captured the directional influence imposed by the model, and the frequency dependent GC matches the one computed with the non-parametric method in Fig.~3.

\subsection{Wilson's algorithm}\label{sec:wilson}
\renewcommand{\thefigure}{E\arabic{figure}}
\renewcommand{\theequation}{E\arabic{equation}}
\renewcommand{\thetable}{E\arabic{table}}
\setcounter{figure}{0}
\setcounter{equation}{0}
\setcounter{table}{0}

This algorithm is used to calculate
the transfer matrix and the covariance matrix used to generate the spectral matrix ${\pmb S}(\omega)$.
It relies on the fact that each of these matrices can be decomposed in terms of basis functions
$\psi(e^{i\omega})$ to be determined recursively.
Once these functions are determined, the $\pmb H$ and $\pmb \Sigma$ matrices that generate the spectral matrices can be inferred.

First, we will briefly describe the mathematical details that allow this process, and then we give a pseudo-code algorithm for those willing to implement the decomposition.
Further mathematical details are given
in the original paper by Wilson~\cite{wilson1972}.

We call the square matrix with dimensions $[m,m]$, the spectral function $\pmb S(\omega)$, defined in the range: $-\pi \leq \omega \leq \pi$.

The spectral function can be written, through the Wiener-Khinchin theorem \cite{laing2010stochastic}, as a function of its correlation coefficients $R_{k}$, as:

\begin{equation}
    \pmb S(\omega) = \sum_{k=-\infty}^{\infty}R_{k}e^{ik\omega},
\end{equation}

\noindent the correlation coefficients can be obtained via the inverse Fourier transform:

\begin{equation}
    R_k = \frac{1}{2\pi}\int_{-\pi}^{\pi}\pmb S(\omega)e^{-ik\omega}d\omega.
    \label{eq:inv_coeff}
\end{equation}

The Wilson's factorization  theorem \cite{wilson1972} states  that the spectral matrix $\pmb S(\omega)$ can be written as the product:

\begin{equation}
    \pmb S(\omega) = \psi(e^{i\omega})\psi(e^{i\omega})^{\dag},
    \label{eq:wilson_theorem}
\end{equation}

\noindent where $\psi$ is a generative function, represented as a square matrix with the same dimensions of $\pmb S$, and can be written as a Fourier series:

\begin{equation}
    \psi(e^{i\omega}) = \sum_{k=0}^{\infty} \pmb A_{k}e^{ik\omega},
\end{equation}

\noindent where $\pmb A_{k}$ are moving average coefficients, and can be obtained via the inverse Fourier transform:

\begin{equation}
    \pmb A_{k} = \frac{1}{2\pi}\int_{-\pi}^{\pi}\psi(e^{i\omega})e^{-ik\omega}d\omega.
    \label{eq:ma_coeff}
\end{equation}

The function $\psi(e^{i\omega})$ can be extended to the unit circle by making $z^{k}=e^{ik\omega}$ ($|z| < 1$) \cite{wilson1972,dhamala2008estimating}, as:

\begin{equation}
    \psi(z) = \sum_{k=0}^{\infty}\pmb A_{k}z^{k},
\end{equation}

\noindent where $\psi(0) = \pmb A_{0}$. The spectral matrix $\pmb S$, and the transfer matrix $\pmb H$ equation~\eqref{eq:spec_mat_def}, can also be represented in the unit circle, with $\pmb H(0) = \pmb I$ \cite{dhamala2008estimating}. Knowing, that we can write equation~\eqref{eq:wilson_theorem}, as:

\begin{equation}
\begin{split}
    \pmb S(z) &= \psi(z)\psi(z)^{\dag}  =\\ &=\sum_{k=0}^{\infty}\pmb A_{k}z^{k}\sum_{k^{'}=0}^{\infty}\pmb A_{k^{'}}^{T}z^{-k^{'}}\\
    &=\sum_{k=0}^{\infty}\sum_{k^{'}=0}^{\infty}\pmb A_{k}\pmb A_{k}^{T}z^{k-k^{'}}.
\end{split}
\label{eq:wilson2}
\end{equation}

Comparing equation~(21) and equation~\eqref{eq:wilson2}, knowing that $\pmb I \pmb M \pmb I^{T} = \pmb M$, where $\pmb M$ is any  square matrix, for $z = 0$, we have:

\begin{equation}
\begin{split}
        \pmb S(0) &= \pmb H(0)\pmb\Sigma\pmb H(0)^{\dag} = \pmb I\pmb\Sigma\pmb I^{\dag} \\&=\pmb \Sigma = \pmb A_{0}\pmb A_{0}^{T},
\end{split}
\end{equation}

\noindent therefore,

\begin{equation}
    \pmb \Sigma = \pmb{A}_{0}\pmb A_{0}^{T}.
    \label{eq:sigma_wilson}
\end{equation}

Further, noticing that $\pmb A_{0}^{-1}\pmb A_{0}\pmb A_{0}^{T}\pmb A_{0}^{-T} = \pmb I$, we may rewrite equation~\eqref{eq:wilson_theorem} as (from here on we omit the function arguments to simplify the notation):

\begin{equation}
    \pmb S = \psi\psi^{\dagger} = \psi\pmb A_{0}^{-1}\underbrace{\pmb A_{0}\pmb A_{0}^{T}}_{\pmb\Sigma}\pmb A_{0}^{-T}\psi^{\dagger}.
    \label{eq:wilson3}
\end{equation}

\noindent Comparing equation~(21) to equation~\eqref{eq:wilson3}, leads to:

\begin{equation}
    \pmb H = \psi\pmb A_{0}^{-1}.
    \label{eq:tm_wilson}
\end{equation}

Before going to the proper Wilson's algorithm, let us define the plus operator $[.]^{+}$ as such that, given a function $g(\omega)$:

\begin{equation}
    g(\omega) = \sum_{k=-\infty}^{\infty}\beta_{k}e^{ik\omega },
    \label{eq:g_func}
\end{equation}

\noindent  the plus operator would be such that:

\begin{equation}
    [g(\omega)]^{+} = 0.5\beta_{0} +  \sum_{k=1}^{\infty}\beta_{k}e^{ik\omega }.
    \label{eq:plus_op}
\end{equation}

Wilson's algorithm consists of finding a solution to equation~\eqref{eq:wilson_theorem}:

\begin{equation}
Y(\omega)=\psi(e^{i\omega})\psi(e^{i\omega})^{\dag}-\pmb S(\omega)=0.
\end{equation}

The function $\psi$, can be found through the iterative solution of equation~\eqref{eq:wilson_iter} (for further details on how to derive this iterative equation see Section~2 in the original paper by Wilson \cite{wilson1972}):
\begin{equation}
    \psi_{k+1} = \psi_{k}\underbrace{\left[\psi_{k}^{-1}\pmb S\left(\psi_{k}^{-1}\right)^\dag + \pmb I\right]^{+}}_{[g]^{+}}
    \label{eq:wilson_iter}
\end{equation}

Below we present a pseudo-code for the Wilson's algorithm, where \textbf{FFT} and \textbf{IFFT} are the fast Fourier transform and its inverse, respectively. \textbf{CHOLE} is the Cholesky factorization and \textbf{TRIU} \cite{burden1980numerical} returns the upper-triangle of a matrix, \textbf{NORM} returns the Euclidean norm of a vector. The \textbf{PlusOperator} algorithm is also presented below where $X[:]$ means that we are getting all values in an specific axis of an array.

\begin{algorithm}
  \caption{Plus operator algorithm. Even though $g$ may be any function, this algorithm is specific for the spectral decomposition discussed here, hence we calculate it for each element of $g$ defined in equation~\eqref{eq:wilson_iter}.}\label{euclid}
  \begin{algorithmic}[1]
  \scriptsize
    \Procedure{PlusOpertor}{$g$}
      \State $m\gets\text{Size of g}(\omega)$
      \State $n\omega\gets\text{Size of }\omega$
      
      \State $\beta\gets\text{zeros}(m,m,n\omega)$ \mycomment{Creating matrix $\beta$}
      \For{\texttt{$i$ from $0$ to  $m$ with step $1$ }}
            \For{\texttt{$j$ from $0$ to  $m$ with step $1$ }}
                    \State $\beta[i,j,:] \gets \text{IFFT}(g[i,j,:])$ \mycomment{Getting coefficients $\beta$ in equation~\eqref{eq:g_func}, via inverse Fourier Transform}
            \EndFor
      \EndFor
      
      \State $\beta_{p} \gets \beta$
      \State $\beta_{0}\gets 0.5*\beta[:,:,0]$   
      \State $\beta_{p}[:,:,0] \gets \text{TRIU}(\beta_{0})$
      \State $\beta_{p}[:,:,n\omega/2:] \gets 0$ \mycomment{Getting only half series as in equation~\eqref{eq:plus_op}}
      
      \State $g_{p}\gets\text{zeros}(m,m,n\omega)$
      
      \For{\texttt{$i$ from $0$ to  $m$ with step $1$ }}
            \For{\texttt{$j$ from $0$ to  $m$ with step $1$ }}
               \State $g_{p}[i,j,:] \gets \text{FFT}(\beta_{p}[i,j,:])$ \mycomment{Returning to the frequency domain to get equation~\eqref{eq:plus_op}}
            \EndFor
      \EndFor
     
     \State \textbf{return} $g_{p}$
    \EndProcedure
  \end{algorithmic}
\end{algorithm}

\begin{algorithm}
  \caption{Wilson’s algorithm. This algorithm relies on the fact the the spectral matrix is Hermitian and can be decomposed in a basis of functions $\psi(e^{i\omega})$, further allowing
  the spectral matrix ${\pmb S}(\omega)$ to be
  decomposed in terms of the product between the transfer matrix and the
  covariance matrix. It first determines the basis recursively, starting from a triangular real Hermitian matrix (calculated by the Cholesky decomposition of the initial correlation matrix). Then, it proceeds
  to calculate the matrices that compose ${\pmb S}(\omega)$.}\label{euclid2}
  \begin{algorithmic}[1]
  \scriptsize
    \Procedure{Wilson}{$S,N_{\rm iter}, \text{tol}$}\mycomment{$S$ is the spectral matrix, $N_{\rm iter}$ the number of iterations, and tol the maximum error tolerance.}
      \State $m\gets\text{Size of S}(\omega)$
      \State $n\omega\gets\text{Size of }\omega$
      
      \State $R\gets\text{zeros}(m,m,n\omega)$ \mycomment{Creating matrix $R$ with auto-correlation values}
      \For{\texttt{$i$ from $0$ to  $m$ with step $1$ }}
        \For{\texttt{$j$ from $0$ to  $m$ with step $1$ }}
            \State \texttt{$R[i,j,:] \gets IFFT(S[i,j,:])$}\mycomment{Computing auto-correlation coeff. (equation~\eqref{eq:inv_coeff})}
        \EndFor
    \EndFor
    \State $\psi\gets\text{zeros}(m,m,n\omega)$ \mycomment{Creating matrix $\psi$, all values set as zero}
    
    \For{\texttt{$i$ from $0$ to  $n_{\omega}$ with step $1$ }}
    \State
    $\psi[:,:,i]\gets\text{CHOLE}(R[:,:,0])$\mycomment{Initial value for $\psi$ is the Cholesky decomposition of $R[:,:,0]$}
    \EndFor
    
    \State $g\gets\text{zeros}(m,m,n\omega)$ \mycomment{Creating matrix to store $g$ (Term indicated by the underbrace in equation~\eqref{eq:wilson_iter})}
    
    \State $I\gets\text{eye}(m,m)$ \mycomment{Creating identity matrix}
    
    \For{\texttt{$t$ from $0$ to  $N_{\rm iter}$ with step $1$ }}\mycomment{Iterate to compute $\psi$ (equation~\eqref{eq:wilson_iter})}
        \For{\texttt{$i$ from $0$ to  $n\omega$ with step $1$ }}
            \State $g[:,:,i] \gets \psi[:,:,i]^{-1}*S[:,:,i]*\left(\psi[:,:,i]^{-1}\right)^{\dagger}$
            \State $g[:,:,i] \gets g[:,:,i] + I$
        \EndFor
        
        \State $g_p = \text{PlusOperator}(g)$ \mycomment{Applying the plus operator (equation~\eqref{eq:plus_op}) in $g$}  
        \State $\psi_{\rm old} \gets \psi$ \mycomment{Saving old value of $\psi$ before updating}
        \State $\psi_{\rm err} \gets 0$ \mycomment{Variable to store the error}
        \For{\texttt{$i$ from $0$ to  $n\omega$ with step $1$ }}\mycomment{Updating $\psi$}
            \State $\psi[:,:,i] \gets \psi[:,:,i]*g_p[:,:,i]$
            \State $\psi_{\rm err} \gets \psi_{\rm err} + \text{NORM}(\psi[:,:,i]-\psi_{\rm old}[:,:,i])/n\omega$
        \EndFor
        
    \If {$\psi_{\rm err} < \text{tol}$}
        \State $break$
    \EndIf
    \EndFor 
    
    \State $A\gets\text{zeros}(m,m,n\omega)$ \mycomment{Creating matrix to store the moving average coeff.}
    
    \For{\texttt{$i$ from $0$ to  $m$ with step $1$ }}
         \For{\texttt{$j$ from $0$ to  $m$ with step $1$ }}
            \State $A[i,j,:] \gets \text{IFFT}(\psi[i,j,:])$\mycomment{Computing the moving average coeff. (equation~\eqref{eq:ma_coeff})}
        \EndFor
    \EndFor
    
    \State $A_{0}\gets A[:,:,0]$
    
    \State $\Sigma \gets A_{0}*A_{0}^{T}$ \mycomment{Computing covariance matrix equation~\eqref{eq:sigma_wilson}}
    
    \State $H\gets\text{zeros}(m,m,n\omega)$ \mycomment{Creating variable to store transfer matrix.}
    
    \For{ \texttt{$i$ from $0$ to  $n\omega$ with step $1$ } }
            \State $H[:,:,i] \gets \psi[:,:,i]*A_{0}^{-1}$ \mycomment{Computing transfer matrix equation~\eqref{eq:tm_wilson}}
    \EndFor
    
  \State \textbf{return} $H, \Sigma$
      
    \EndProcedure
  \end{algorithmic}
\end{algorithm}

\end{document}